\documentclass[aps,preprint]{revtex4}
\usepackage{amsfonts}
\usepackage{amsmath}
\usepackage{amssymb}
\usepackage{graphicx}
\usepackage{epstopdf}
\usepackage{epsfig}
\usepackage{amsfonts,amsthm,eucal,amssymb,amscd}

\begin{document}

\title{The hyperbolic step potential: Antibound states, SUSY partners and Wigner time delays}
\author{M Gadella$^1$, S Kuru$^2$, J Negro$^1$ }
\affiliation{$^1$Departamento de F\'isica Te\'orica, At\'omica y Optica and IMUVA, Universidad de Valladolid, E-47011, Valladolid, Spain}
\affiliation{$^2$Department of Physics, Faculty of Science, Ankara University, 06100 Ankara, Turkey}

\begin{abstract}
 We study the scattering produced by a one dimensional hyperbolic step potential, which is exactly solvable and shows an unusual interest because of its asymmetric character. The analytic continuation of the scattering matrix in the momentum representation has 
 a branch cut and an infinite number of simple poles on the negative imaginary axis which are related with the so called anti-bound states. This model does not show resonances. Using the wave functions of the anti-bound states, we obtain supersymmetric (SUSY) partners  which are the series of Rosen Morse II potentials. We have computed the Wigner reflection and transmission time delays for the hyperbolic step and such SUSY partners.
Our results show that the more bound states a partner  Hamiltonian has the smaller is the time delay.  We also have evaluated time delays for the hyperbolic step potential in the classical case and have obtained striking similitudes with the quantum case. 
\end{abstract}

\maketitle

\section{Introduction}

One dimensional models in quantum mechanics are important because they are much more tractable and sometimes they are solvable, what makes them quite suitable for testing many properties and  the behavior of a wide range of quantum systems. In particular, we may analyze the properties of quantum states. In non-relativistic quantum mechanics, we usually deal with four types of states: bound states, scattering states, resonance states and anti-bound also called virtual states. The present article focus its attention on the latter. 

Since resonances and anti-bound states appear in resonance scattering, it is customary to use a characterization of both that relies on the properties  of the analytic continuation into the complex plane of the scattering matrix ($S$-matrix) either in the momentum, ${\cal S}(k)$, or in the energy representation, ${\cal S}(E)$. In the sequel, we shall restrict ourselves to the momentum representation. While resonances are given by poles of any multiplicity located in the lower half plane symmetrically with respect to the imaginary axis, anti-bound states are given by simple poles on the negative imaginary axis \cite{NU,BO}. Resonances and anti-bound states are particular cases of quantum transients \cite{GC}.

Physically, anti-bound states have been indirectly observed at low energy when scattering shows an anomalous large cross section \cite{NU}.  Anti-bound states may also produce a long time delay, as shown in \cite{BO}, Chapter XVIII. A typical example is the anti-bound state produced in the scattering neutron-proton \cite{BW}, where a anti-bound of deuteron is produced. Anti-bound states are also observed in nuclear physics. In this context,  they usually manifest themselves by the capture of a neutron by some light nuclei, for instance, $^{10}$Li which is a typical example or $^9$Be \cite{JAP}. 
Some further discussions on the physical properties of anti-bound states may be found in \cite{NA}.

In the present paper, we show that the hyperbolic step potential has anti-bound states. Because of its shape, this potential may be used as an approximation of the  Woods-Saxon potential, which is a common device in different studies in nuclear physics \cite{WS,CDNSW,LNPD,SWV}. Studies on the shell model with anti-bound states produced by the Woods-Saxon potential have appeared in the literature \cite{BV,ROLO,DRSL,SBV}.  Nevertheless, there is an important difference: The Woods-Saxon potential is accompanied by an impenetrable barrier at the origin, while the hyperbolic step potential is a genuine one dimensional potential defined for all values of $x$. Consequently, the hyperbolic barrier potential is similar to a square barrier. In particular, it does not have bound states. Instead, it has a rich structure of anti-bound states and this makes it particularly attractive. 

Although experimentally observable anti-bound states are of low negative energy, in specific models poles of the analytic continuation of the S-matrix  on the negative imaginary axis may have any negative energy value.  This is the case of the one dimensional semi-oscillator with a contact potential at the origin \cite{AGLM} or in the one dimensional hyperbolic P\"oschl-Teller potential \cite{CGKN}. 

In the energy representation, anti-bound states are represented for poles in the analytic continuation of the $S$-matrix on the negative real axis in the second sheet of the two sheeted Riemann surface corresponding to the transformation $k=\sqrt{2mE/\hbar}$. As is the case with Gamow states for resonances, anti-bound states could be represented by non-normalizable eigenfunctions of the Hamiltonian with negative real eigenvalues given by the real poles of the ${\cal S}(E)$ matrix  on the second sheet \cite{GA}.  

Many models for resonance scattering showing resonances and anti-bound states require numerical approximations in order to find poles of ${\cal S}(k)$, or equivalently of ${\cal S}(E)$. In order to study relevant properties of these physical states, it would be important to find exactly solvable models for resonances and anti-bounds. In a recent paper \cite{CGKN}, we have shown that this is the case for the hyperbolic P\"oschl-Teller potential. This exact solvability has permitted to show the existence of ladder operators connecting series of bound {and} anti-bound states as well as series of resonance states. The same is suspected to happen for a range of potentials of hypergeometric type  \cite{KC,SH,RM}. In addition exact solvability may be used for another purposes, like for instance to check the accuracy of numerical computational methods \cite{AGLM,GGL}. 

The hyperbolic step model offers quite interesting features because of its asymmetric character. As a consequence of this asymmetry, the matrix ${\cal S}(k)$ is not unitary, but instead it satisfies a somehow modified unitarity relation, as we shall see. In this case, the analytic continuation of the matrix ${\cal S}(k)$ has a branch cut, which is not present for  
symmetric asymptotic conditions. 
Another outcome of the asymmetry is that this potential has two different values of the momentum $k$ and $k'$  at asymptotic regions left and right, respectively.  This implies, for instance, that in order to have bound (or anti-bound) states the corresponding conditions should be satisfied in both asymptotic regions:
$k$ and $k'$ must be in the positive imaginary  (or negative imaginary) axis.

Resonance and anti-bound states have been used for the construction of supersymmetric partners of a given Hamiltonian \cite{CGKN,oscar1,oscar2}.  In general, to obtain a SUSY partner from a given initial potential, one uses eigenfunctions of such Hamiltonian  having specific properties  \cite{CKS,CHIS,FLISHFLASH,MO}. In our case, we will use  just the wave functions of the anti-bound states in order to obtain a hierarchy of SUSY partners of the hyperbolic step potential  called Rosen Morse II potentials.

After interacting with a potential in a collision process and being partially reflected and transmitted, a wave packet undergoes a time delay with respect to the time employed by the free motion. There are some methods to measure this time, one of the most common relies in the definition of the Wigner time delay \cite{BO,W,S,T}.   We have calculated the Wigner time delay for the hyperbolic step Hamiltonian and some of its Rosen Morse II partners.  The absence of bound states for this Hamiltonian and the existence of bound states for its partners affects the phase and time delay that can be measured.  We have also compared these time delays with those obtained for a classical hyperbolic step potential.  

After this presentation, we summarize the organization of this article as follows. In Section 2, we  solve the Schr\"odinger equation for the hyperbolic step potential and obtain the asymptotic solutions. In Section 3, we analyze  scattering properties of this potential, in particular we compute the anti-bound poles and states. We see that the asymptotic asymmetry of the potential leads to new analytic properties of the $S$-matrix elements. Section 4 is devoted to the application of the factorization method to the hyperbolic step potential in order to find its SUSY partners derived from the eigenfunctions corresponding to the anti-bound states. Next, in Section 5, we compute the phases and time delays for the hyperbolic
step potential as well as for some of its partner potentials. These values can be
computed analytically.  Finally, the paper ends with some conclusions and remarks.

\section{The hyperbolic step potential}

Our first objective is an analysis of the Hamiltonian with a hyperbolic step potential. To begin with, we find a solution in terms of  hypergeometric functions. We  also compute the scattering matrix in order to investigate the possible existence of bound and anti-bound states as well as resonances. 

\begin{figure}
\centering
\includegraphics[width=0.40\textwidth]{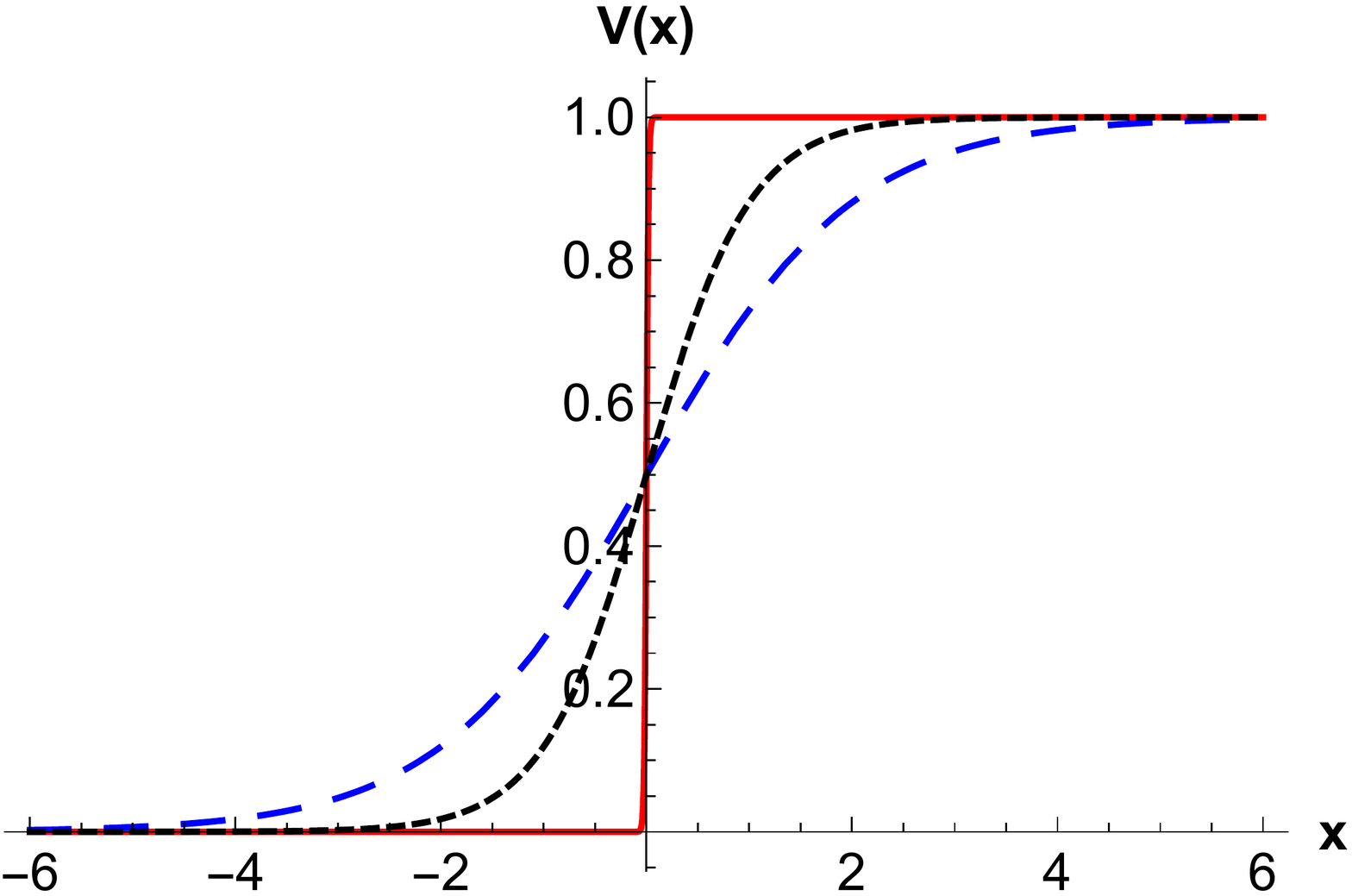}\,\qquad
\includegraphics[width=0.40\textwidth]{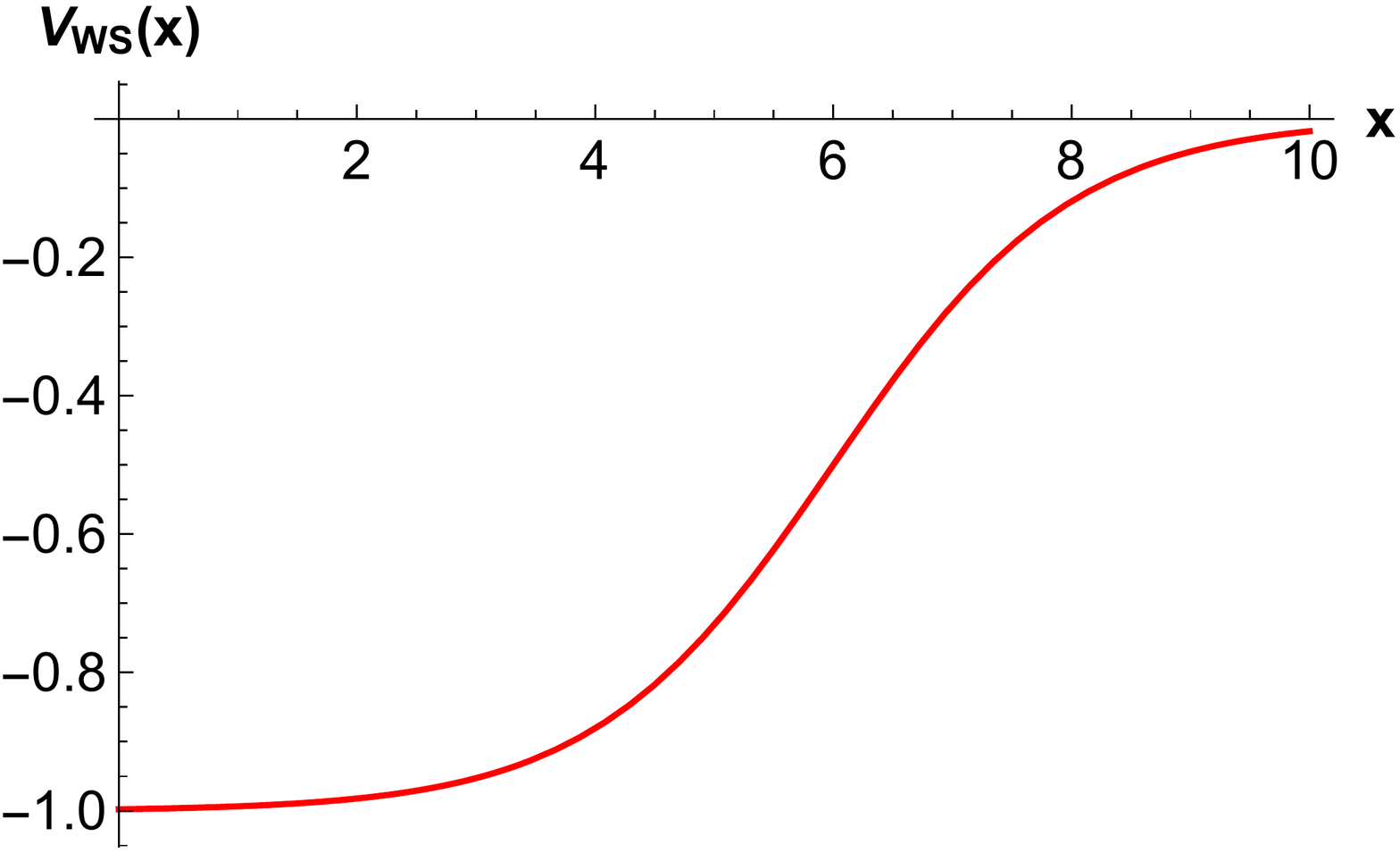}
\caption{\small At the left: Plot of the hyperbolic step potential for $V_0=1$ and different
values of $\alpha$. The continuous line corresponds to
$\alpha=0.001$, the dotted line to $\alpha=0.5$, the dashed line to
$\alpha=1$. At the right: Plot of the Woods-Saxon potential, $V_{{\rm WS}}(x)=\dfrac{1}{2}(\tanh[\dfrac{ x }{2}-3]-1)$.}\label{pot}
\end{figure}

The one dimensional hyperbolic step potential has the following form:
\begin{equation}\label{1}
V(x)=\frac{1}{2} V_0\,\left(1+ \tanh\frac{x}{2\,\alpha}\right)\,, \qquad V_0>0\,,\;\alpha>0\,.
\end{equation}
The parameter $V_0$ is the barrier height. In order to figure out the shape of $V(x)$ and to compare it with Woods-Saxon potential, we have plotted both of them in Figure \ref{pot} for $V_0=1$ and three different values of $\alpha$. When $\alpha \to 0$ this potential goes into
the step potential.

The Hamiltonian for this potential is
\begin{equation}\label{2}
H=-\frac{\hbar^2}{2m}\,\frac{d^2}{dx^2}+ \frac{1}{2} V_0\,\left(1+
\tanh\frac{x}{2\,\alpha}\right)\,.
\end{equation}
The stationary Schr\"odinger equation for this Hamiltonian has already been solved. See for instance  \cite{FLU}. Nevertheless, we recall this solution here to guide the reader,  for the sake of completeness and for a proper introduction of the notation to be used in the sequel. The Schr\"odinger equation reads:
\begin{equation}\label{3}
\psi''(x)+\left[ k^2-\frac{m\,V_0}{\hbar^2} \left( 1+\tanh \frac{x}{2\alpha}  \right)  \right]\psi(x)=0\,,
\end{equation}
where $k^2=(2mE)/\hbar^2$ and $\psi''(x)$ is the second derivative with respect to the variable $x$ of the wave function $\psi(x)$. In order to find solutions of (\ref{3}), we use a new variable $y$ defined through its dependence on $x$ as
\begin{equation}\label{4}
y(x):=(1+e^{x/{\alpha}})^{-1}\,.
\end{equation}
Then, $0<y< 1$ and $x=\alpha\log(y^{-1}-1)$. Thus, if we define $U(y):= \psi(\alpha\log(y^{-1}-1))$, equation (\ref{3}) becomes
\begin{equation}\label{5}
y(1-y)\,U''(y)+(1-2y)\,U'(y)+\left[\frac{\kappa^2}{(1-y)y}-\frac{\lambda^2}{y}\right]U(y)=0\,,
\end{equation}
with
\begin{equation}\label{6}
\lambda^2=\dfrac{2mV_0}{\hbar^2}\alpha^2\,,\qquad
\kappa^2=\alpha^2\,k^2\,.
\end{equation}
In order to transform (\ref{5}) into a hypergeometric equation, we need to introduce the new indeterminate $f(y)$, defined as
\begin{equation}\label{7}
f(y):= U(y)\,y^{-\nu}\,(1-y)^{-\mu}\,,
\end{equation} 
where, $\nu^2=\lambda^2-\kappa^2$ and $\mu^2=-\kappa^2$.  Let us use (\ref{7}) into (\ref{5}) so as to obtain
\begin{equation}\label{8}
y(1-y)\,f''(y)+[2\nu+1-y(2\mu+2\nu+2)]\,f'(y)-[(\mu+\nu)(\mu+\nu+1)]\,f(y)=0\,.
\end{equation}
This equation has the form of a  hypergeometric  differential equation, which has the form 
$
y(1-y)\,f''(y)+[c-(a+b+1)y]\,f'(y)-ab\,f(y)=0\,
$. As far as the constant $c$ is not an integer, its general solution is given by a linear combination of Kummer, also called hypergeometric, functions of the form   
 \cite{AS}:
$
f(y)=C\,_2F_1(a,b;c;y)+D\, y^{1-c}\,_2F_1(1+a-c,1+b-c; 2-c;y)\,,
$ 
where $C$ and $D$ are arbitrary constants.  
Comparing the equation (\ref{8}) with the hypergeometric equation, we obtain    
$a=\nu+\mu$, $b=\nu+\mu+1$ and $c=1+2\nu$.  Finally, we can reverse transformations (\ref{7}) and (\ref{4}) so that the general solution of equation (\ref{3}) is given by

\begin{eqnarray}\label{9}
\psi(x)= \frac C4\, \left(1+\tanh{\frac{x}{2\,\alpha}}\right)^{i\,\alpha\,k}\left(1-\tanh{\frac{x}{2\,\alpha}}\right)^{-i\,\alpha\,k'} \nonumber\\[2ex] \times\, _{2}
F_{1}\left(\mu+\nu,\mu+\nu+1;1+2\nu;\frac{1}{2}\left(1-\tanh{\frac{x}{2\,\alpha}}\right)\right)  \nonumber\\[2ex] +\frac D4\, \,\left(1+\tanh{\frac{x}{2\,\alpha}}\right)^
{i\,\alpha\,k}\left(1-\tanh{\frac{x}{2\,\alpha}}\right)^{-i\,\alpha\,k'}  \nonumber\\[2ex]  \times\,  _{2}F_{1}\left(\mu-\nu,\mu-\nu+1;1-2\nu;\frac{1}{2}\left(1-\tanh{\frac{x}{2\,\alpha}}\right)\right)\,,
\end{eqnarray}
where 

\begin{eqnarray}\label{10}
k'=\sqrt{\dfrac{2m(E-V_0)}{\hbar^2}}=\sqrt{k^2-\dfrac{\lambda^2}{\alpha^2}}\,,\qquad \nu=-i\,\alpha\,k'\,,\qquad \mu=i\,\alpha\,k \,.
\end{eqnarray}

Once we have found the general solution for (\ref{3}), our next goal is the analysis of the transfer and scattering matrices.

\subsection{The scattering and transfer matrices}

The scattering matrix ${\mathcal S}(k)$  connects the asymptotic forms of the incoming wave
function with the outgoing wave function. Using the asymptotic behavior
of the hypergeometric functions \cite{AS}, we find the asymptotic forms
of the solution (\ref{11}) to the left, $x\to -\infty$, and to the right, $x\to+\infty$, which are

\begin{itemize}
\item
For $x\to-\infty$
\begin{equation}\label{11}
\begin{array}{rl}
\!\!\!\!\!\!\!\!\!\!\!\!\!\!\!\!\!\!\!\psi^-(x)&= \displaystyle \left[
C\, \frac{\Gamma\left(1+2\nu\right)\,\Gamma\left(-2\mu\right)}
{\Gamma\left(-\mu+\nu+1\right)\, \Gamma\left(-\mu+\nu\right)} +D\,
\frac{\Gamma\left(1-2\nu\right)\,\Gamma\left(-2\mu\right)}
{\Gamma\left(-\mu-\nu+1\right)\, \Gamma\left(-\mu-\nu\right)}
\right]\,e^{ikx}
\\[2.75ex]
&+ \displaystyle\left[C\,
\frac{\Gamma\left(1+2\nu\right)\,\Gamma\left(2\mu\right)}
{\Gamma\left(\mu+\nu\right)\, \Gamma\left(\mu+\nu+1\right)} +D\,
\frac{\Gamma\left(1-2\nu\right)\,\Gamma\left(2\mu\right)}
{\Gamma\left(\mu-\nu\right)\, \Gamma\left(\mu-\nu+1\right)}
\right]\,e^{-ikx}\,
\\[3ex]
&=\displaystyle A\,e^{ikx}+B\,e^{-ikx}\,.
\end{array}
\end{equation}

\item
For $x\to+\infty$
\begin{equation}\label{12}
\!\!\!\!\!\!\!\!\!\!\!\!\!\!\!\!\!\!\!\psi^+(x)=
C\,e^{ik'x}+D\,e^{-ik'x}\, .
\end{equation}
\end{itemize}
Here, $\Gamma(z)$ is the Euler Gamma function \cite{AS}.  We recall that $k$ and $k'$ correspond to the momentum at the asymptotic regions $x \to +\infty$ and $x \to -\infty$, respectively.

Then, $\mathcal S(k)$ will relate the amplitudes given by (\ref{11}) and (\ref{12}), respectively. One may write this relation as \cite{BOYA}
\begin{equation}\label{13}
\left( \begin{array}{c} B \\[2ex] C        \end{array}   \right) = \left( \begin{array}{cc}
S_{11}  &  S_{12}  \\[2ex]  S_{21}  &  S_{22}    \end{array}   \right) \left( \begin{array}{c} A \\[2ex] D
\end{array}   \right)\,,
\end{equation}
where the scattering matrix $\mathcal S(k)$ is the $2\times 2$ matrix with entries $S_{ij}$.  On the other hand, the transfer matrix $\mathcal T(k)$,  relates the amplitudes of the asymptotic wave functions in $x\to -\infty$, with the ones in  $x\to+ \infty$.  This is

\begin{equation}\label{14}
\left(\begin{array}{c}
C \\[2ex]
D\end{array}\right) =
\left(\begin{array}{cc}
T_{11} & T_{12} \\[2ex]
T_{21} & T_{22}
\end{array}
\right) \left(\begin{array}{c}
A \\[2ex]
B\end{array}\right)\,.
\end{equation}

The  entries $T_{ij}$ of the transfer matrix have  a simple relation with those of the scattering matrix   $S_{ij}$ given by

\begin{equation}\label{15}
\mathcal S(k)= \frac{1}{T_{22}}\left(\begin{array}{cc}
-T_{21} & 1 \\[2.ex]
T_{11} T_{22}-T_{21}T_{12} & T_{12}
\end{array}
\right)\,.
\end{equation}
The explicit form of the transfer matrix $\mathcal T(k)$ can be readily obtained from (\ref{11}) and (\ref{12}):

\begin{equation}\label{16}
\mathcal T(k)=\frac{k}{k'}\left(\begin{array}{cc}
\dfrac{\Gamma\left(2 i \,\alpha\,k\right)\,\Gamma\left(1+2 i \,\alpha\,k'\right)}
{\Gamma\left(i\,\alpha\,(k+k')\right)\, \Gamma\left(1+i\,\alpha\,(k+k')\right)} &
-\dfrac{\Gamma\left(-2 i \,\alpha\,k\right)\,\Gamma\left(1+2 i \,\alpha\,k'\right)}
{\Gamma\left(-i\,\alpha\,(k-k')\right)\, \Gamma\left(1-i\alpha(k-k')\right)}
\\[2.5ex]
-\dfrac{\Gamma\left(2 i \,\alpha\,k\right)\,\Gamma\left(1-2 i \,\alpha\,k'\right)}
{\Gamma\left(i\,\alpha\,(k-k')\right)\, \Gamma\left(1+i\,\alpha\,(k-k')\right)} &
\dfrac{\Gamma\left(-2 i \,\alpha\,k\right)\,\Gamma\left(1-2 i \,\alpha\,k'\right)}
{\Gamma\left(-i\,\alpha\,(k+k')\right)\, \Gamma\left(1-i\,\alpha\,(k+k')\right)}
\end{array}
\right)\,,
\end{equation}
where $k'$ is given in (\ref{10}).  Using the properties of the Euler Gamma function $\Gamma(z)$ \cite{AS} it can be shown that $\det\mathcal T(k)=k/k'=-\mu/\nu$. 

One important property shows that $\mathcal S(k)$ is not unitary, but instead it satisfies a relation of the type

\begin{equation}\label{17}
\mathcal S^\dagger(k)\,K\,\mathcal S(k)=K\,,\qquad K
=\left(\begin{array}{cc} k & 0\\ 0 & k'     \end{array}   \right)\,,
\end{equation}
where the dagger means the adjoint of the matrix $\mathcal S(k)$. Therefore, $\mathcal S(k)$ is unitary if and only if $k=k'$. 

The current density to the left and to the right asymptotic regions have the following expressions:

\begin{equation}\label{18}
J_L=\frac{\hbar k}{m} (|A|^2-|B|^2),\qquad J_R=\frac{\hbar k'}{m} (|C|^2-|D|^2)\,.
\end{equation}
Taking into account  (\ref{14}) and (\ref{16}), we check that the current is conserved: $J_L=J_R$ (even though $k\ne k'$). This property is consistent with (\ref{17}) and with the self-adjointness of (\ref{2}) \cite{CLMM}. 

By purely outgoing boundary conditions, we mean that only asymptotically outgoing wave functions exist. The wave function $\psi(x)$, solution of (\ref{3}), satisfies these conditions if and only if it shows the following asymptotic form:

\begin{equation}\label{19}
\psi^-(x)= B e^{-i k x}\quad \xleftarrow[\ -\infty\, \leftarrow\, x \  ] \quad \quad\psi(x) 
\quad\xrightarrow[\  x\,\to\,+\infty\  ]\quad\quad \psi^+(x)= C e^{i k' x}\, .
\end{equation}

Equations (\ref{11}) and (\ref{12}) show that this happens if and only if $A=D=0$. From (\ref{14}), we see that $D=T_{21}A+T_{22}B$, so that condition $A=D=0$ implies  $T_{22}(k)=0$. Then, from the expression of
$\mathcal S(k)$ given in  (\ref{15}), we conclude that the solutions of equation  $T_{22}(k)=0$, for $k$ complex, give the poles of $\mathcal S(k)$.  Among such states one can identify bound and anti-bound states as well as resonances. 

According to (\ref{9}),  wave functions
for purely outgoing states are given, up to a constant factor,
by 

\begin{eqnarray}\label{20}
\varphi(x)= \frac{1}{4}\,\left(1+\tanh{\frac{x}{2\,\alpha}}\right)^
{i\,\alpha\,k}\left(1-\tanh{\frac{x}{2\,\alpha}}\right)^{-i\,\alpha\,k'}
\nonumber\\[2ex]   \times\, _{2}F_{1}\left(i \,\alpha\,(k+k'),1+i \,\alpha\,(k+k');1+2 i \,\alpha\,k';\frac{1}{2}\left(1-\tanh{\frac{x}{2\,\alpha}}\right)\right)\,,
\end{eqnarray}
where $k$ is one of the poles of the scattering matrix. Thus, (\ref{20}) represents the wave function for a bound, antibound or resonance state if and only if the pole at $k$ is a bound, antibound or resonance pole of the scattering matrix. We recall on the relation between $k$ and $k'$ given by (\ref{10}). 

\section{Scattering analysis of the hyperbolic step potential}

Our next goal is to obtain reflection and transmission coefficients for real values of $k$. To this end, we solve the equation $T_{22}(k)=0$ for complex $k$, so that we are able to identify the nature of the different types of poles of the scattering matrix. 
For many models, this is a  transcendental equation. However, due to the particular form of the entries of the transfer matrix $\mathcal T(k)$ for the hyperbolic step potential in terms of Gamma functions, this equation is exactly solvable in our case. 

\subsection{Scattering amplitudes}\label{3a}

Scattering is produced for energies above the potential height, i.e., $E>V_0$. An incoming plane wave  from the left, with $E>V_0$, will undergo a reflection as well as a transmission due to the presence of the potential. The asymptotic behavior of the wave function describing this process is given by
\begin{equation}\label{21}
\psi^-(x)= e^{i k x}+ r(k)\,e^{-i k x}\,,\qquad \psi^+(x)= t(k)\, e^{i k' x}\,,
\end{equation}
where $\psi^-(x)$ and $\psi^+(x)$ are the asymptotic forms of $\psi(x)$ when $x\to-\infty$ and $x\to +\infty$, respectively, where the relation between $k$ and $k'$ is given by (\ref{10}). We denote by $r(k)$ the reflection and by $t(k)$ the transmission amplitudes.  In order to obtain $r(k)$ and $t(k)$, we choose $A=1$ and $D=0$ in (\ref{11}) and (\ref{12}). Then, $r(k)=B$ and $t(k)=C$. Using (\ref{14}) and taking (\ref{15}) into account, we have that

\begin{equation}\label{22}
r(k)= B= S_{11} = -\frac{T_{21}}{T_{22}} = \frac{\Gamma\left(2 i \,\alpha\,k\right)\,
\Gamma\left(- i \,\alpha\,(k+k')\right)\,\Gamma\left(1- i \,\alpha\,(k+k')\right)}{\Gamma\left(-2 i \,\alpha\,k\right)\,\Gamma\left(i \,\alpha\,(k-k')\right)\,
\Gamma\left(1+i \,\alpha\,(k-k')\right)}\,,
\end{equation}
and
\begin{eqnarray}\label{23}
t(k)=C=S_{21}= \frac{T_{11}T_{22}-T_{12}T_{21}}{T_{22}}=\frac{k'/k}{T_{22}}=  \frac{\Gamma\left(-i\,\alpha\,(k+k')\right)\, \Gamma\left(1-i\,\alpha\,(k+k')\right)}{\Gamma\left(-2 i \,\alpha\,k\right)\,\Gamma\left(1-2 i \,\alpha\,k'\right)}\,.
\end{eqnarray}

As usual, the reflection $R(k)$ and transmission $T(k)$ coefficients are given by

\begin{equation}\label{24}
R(k)=|r(k)|^2=\left| \dfrac{\Gamma\left(2 i \,\alpha\,k\right)\,
\Gamma\left(- i \,\alpha\,(k+k')\right)\,\Gamma\left(1- i \,\alpha\,(k+k')\right)}{\Gamma\left(-2 i \,\alpha\,k\right)\,\Gamma\left(i \,\alpha\,(k-k')\right)\,
\Gamma\left(1+i \,\alpha\,(k-k')\right)}\right|^2
\,,
\end{equation}
and
\begin{equation}\label{25}
T(k)=|t(k)|^2= \left|
\dfrac{\Gamma\left(-i\,\alpha\,(k+k')\right)\, \Gamma\left(1-i\,\alpha\,(k+k')\right)}{\Gamma\left(-2 i \,\alpha\,k\right)\,\Gamma\left(1-2 i \,\alpha\,k'\right)}\right|^2\,.
\end{equation}

\begin{figure}
\centering
\includegraphics[width=0.45\textwidth]{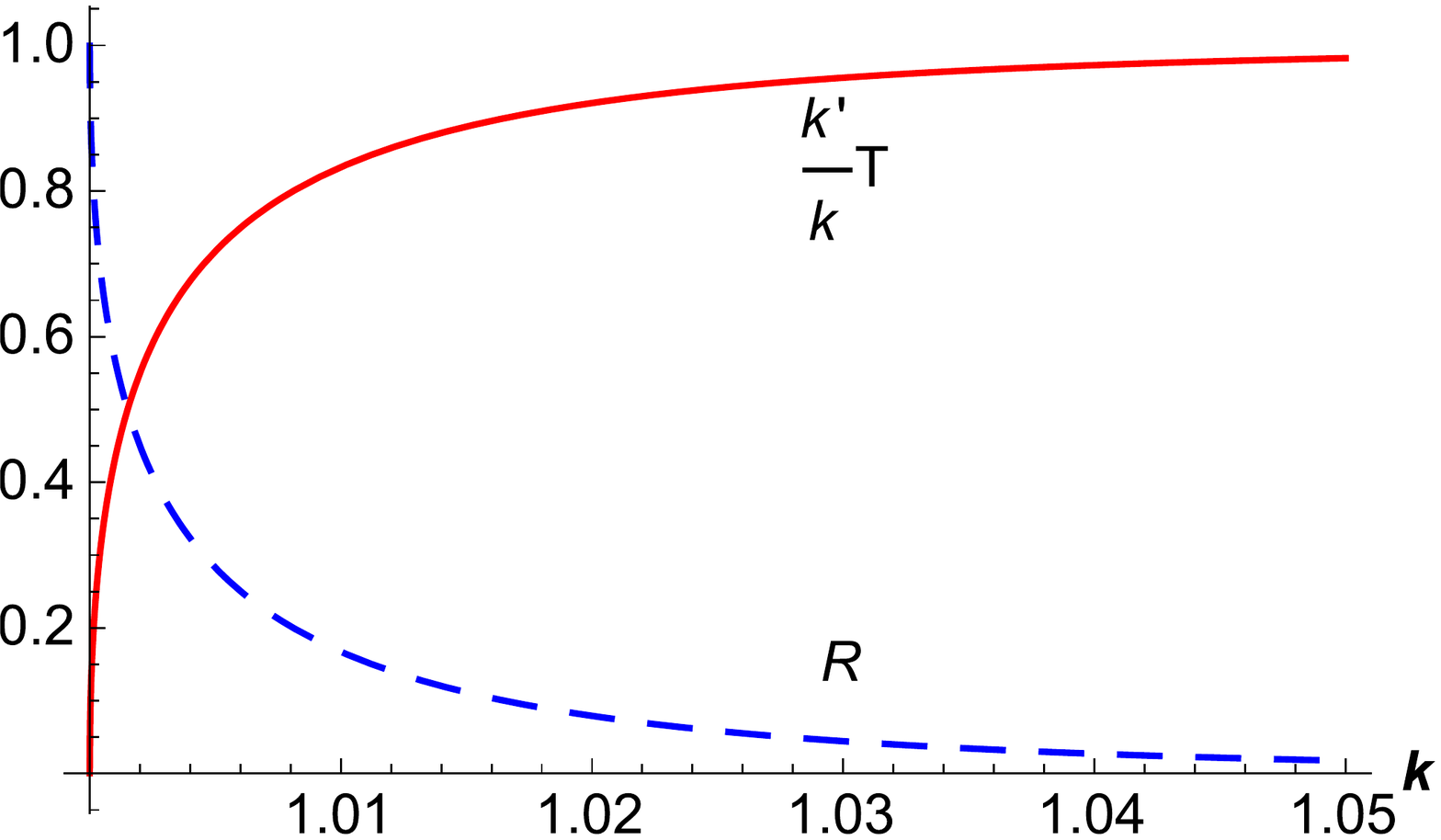}
\caption{\small The continuous line shows the transmission coefficient $T(k)$, while the dashed line represents the reflection coefficient $R(k)$.} \label{fig2}
\end{figure}

We can check that $\dfrac{k'}{k}T(k)+R(k)=1$ for $k^2>2m V_0/\hbar^2$ \cite{GP}.
In Figure \ref{fig2}, we plot the transmission versus the reflection coefficients. As a matter of fact, we are really interested in the explicit expressions for the amplitudes $r(k)$ and $t(k)$, since their phases are related to the delay times that we shall investigate later. 

\begin{figure}
\centering
\includegraphics[width=0.60\textwidth]{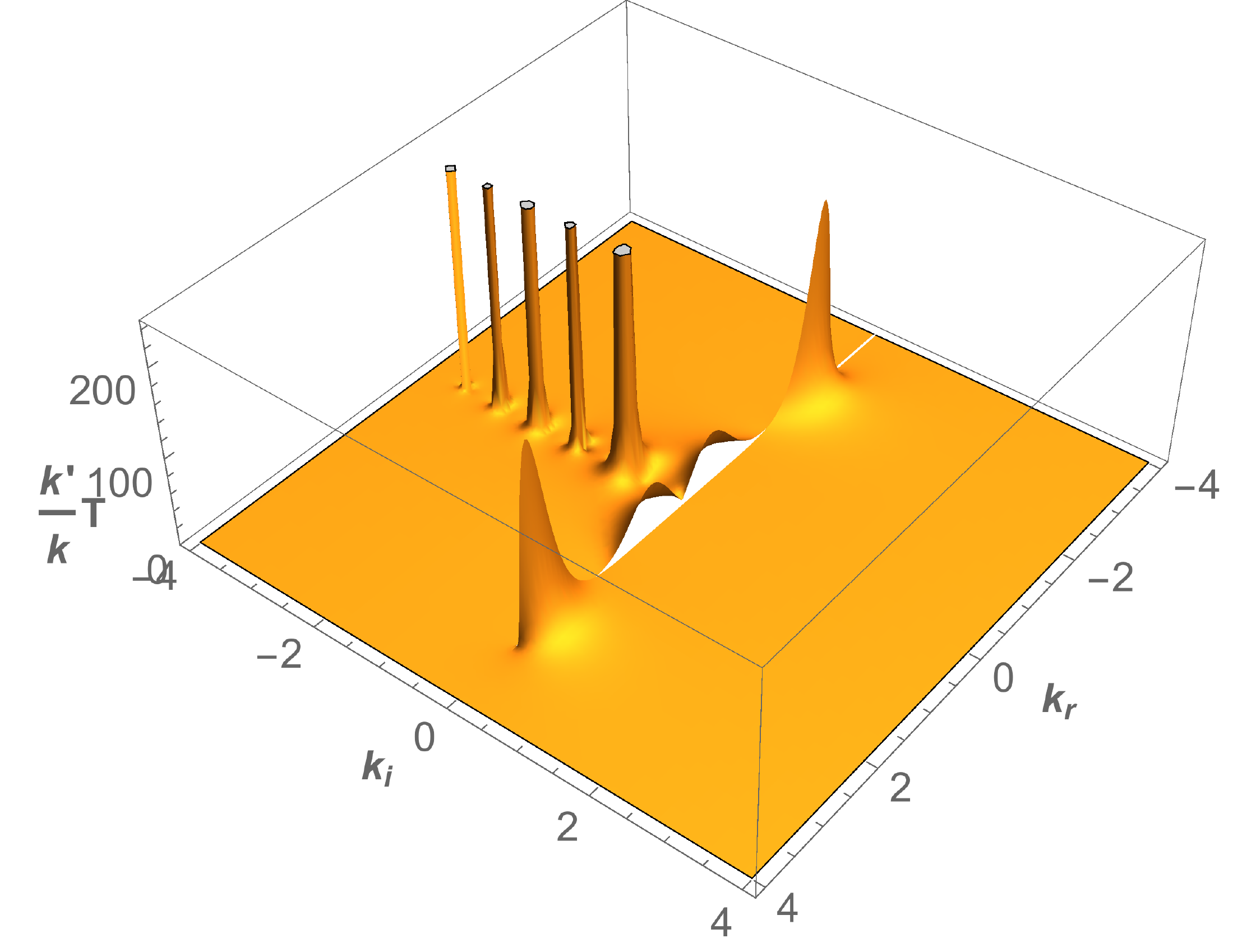}
\caption{\small Plot of $\frac{|k'|}{|k|}\,T(k)$ in terms of $k$ for $V_0=9, \alpha=1$. Here, $k_r$ and $k_i$ are the real and imaginary parts of $k$. Bumps are poles and the white pattern between the real  points $k=-3$ and $k=3$ corresponds to the branch cut. There is no
pole for $k_i>0$. } \label{fig3}
\end{figure}

\subsection{The search for singularities of $\mathcal S(k)$}

Now, we investigate the possible presence of bound, anti-bound and resonance states for the hyperbolic step Hamiltonian (\ref{2}). As was already stated, this may be achieved by searching complex solutions on $k$ of the equation $T_{22}(k)=0$. Therefore, we have to solve the equation

\begin{equation}\label{26}
T_{22}(k)=\dfrac{k}{k'}\,\dfrac{\Gamma\left(-2 i \,\alpha\,k\right)\,\Gamma\left(1-2 i \,\alpha\,k'\right)}
{\Gamma\left(-i\,\alpha\,(k+k')\right)\, \Gamma\left(1-i\,\alpha\,(k+k')\right)}=0\,.
\end{equation} 

Since the Gamma function, $\Gamma(z)$, has no zeroes on the complex plane, solutions of (\ref{26}) are those complex values of $k$ that correspond to poles of the denominator. By comparison between (\ref{23}) and (\ref{26}), we note that these solutions are also the poles of the transmission amplitude $t(k)$, and hence of the transmission coefficient $T(k)$. 
This coefficient  $T(k)$  depends on $k$ both explicitly and also implicitly through $k'$. 

It is clear that $k'(k)$ as a function of the complex variable $k$ given in (\ref{10}), due to the square root, shows a branch cut connecting the branch points $k=\pm\,\lambda/\alpha$. As a consequence the function $T(k)$  also shows a branch cut. This can be seen in  Figure \ref{fig3}. 

As a consequence of the above comments, the poles of $T(k)$  are given by the equation
\begin{equation}\label{27}
-i\,\alpha\,(k+k') =-n, \qquad n=0,1,2,\dots \,.
\end{equation}
From (\ref{27}) and (\ref{10}), we have that $k'=\sqrt{k^2-\lambda^2/\alpha^2}=-(k+in/\alpha)$. Take squares in both sides and perform an obvious simplification. This procedure gives an expression of $k$ as function of $n$, $k(n)$, which is

\begin{equation}\label{28}
k(n)=-\frac{i}{2\,\alpha}\,\left(n-\frac{\lambda^2}{n}\right)\,,
\qquad\qquad n=1,2,\dots\,.
\end{equation}

Taking into account that $\alpha>0$, we may have two possibilities for $k(n)$:

\begin{itemize}

\item{There are some natural values $n$ for which $n-\lambda^2/n<0$. This is possible for large values of $\lambda$. This situation is intriguing, because then equation (\ref{28}) gives positive values for $k(n)$ for a finite number of values of $n$.  If all values of $k(n)$ were zeroes of $T_{22}(k)$ and, hence, poles of the scattering matrix, this would mean that there were poles of the $S$ matrix on the positive imaginary semi-axis. These poles correspond to bound states \cite{NU} for (\ref{2}).  However,  we know that (\ref{2}) does not have bound states.  

The solution to this puzzle lies on the fact that poles of the scattering matrix are given by solutions of (\ref{27}) and not solutions of (\ref{28}). In the transit from (\ref{28}) to (\ref{27}), we have squared an expression. As a result, we have eliminated a square root and this procedure makes (\ref{27}) and (\ref{28}) inequivalent. 

In order to show that $k(n)$, with  $n-\lambda^2/n<0$, cannot be poles of the $S$ matrix, let us write $k(n)=i\beta(n)$. It is obvious that for these values of $k(n)$, $\beta(n)>0$. Then, define $\rho(n):=\{\lambda^2/\alpha^2+\beta^2(n)\}^{1/2}$ and $\theta:= \arg\{\lambda/\alpha+i\beta(n)\}$. The principal branch of the square root is given by the property $-\pi\le \theta<\pi$. Then, consider the identity:

\begin{eqnarray}\label{29}
k'=\sqrt{k^2(n)-\lambda^2/\alpha^2} =\sqrt{i\beta(n)+\lambda/\alpha}\;\sqrt{i\beta(n)-\lambda/\alpha} \nonumber\\[2ex] = \rho(n)\,e^{i\theta/2}\,\rho(n)\,e^{i(\pi/2-\theta/2)} = \rho^2(n)\,e^{i\pi/2}=i\,\rho^2(n)\,.
\end{eqnarray}

If we insert (\ref{29}) in the left hand side of  (\ref{27}), we have

\begin{eqnarray}\label{30}
-i\alpha(k(n)+k'(n))=-i\alpha\left\{ i\beta(n)+i\left\{\beta^2(n)+\frac{\lambda^2}{\alpha^2}\right\}\right\} \nonumber\\[2ex] = \alpha\left\{\beta(n)+\beta^2(n)+\frac{\lambda^2}{\alpha^2}\right\}>0\,,
\end{eqnarray}
which is incompatible with (\ref{27}). Therefore, $k(n)$ with $n-\lambda^2/n<0$ are regular points of the scattering matrix and do not correspond to any particular kind of state. }

\item{The other possible situation gives $n-\lambda^2/n>0$. By repeating the same arguments as above, we conclude that (\ref{27}) is now fulfilled. Consequently, values $k(n)$ in (\ref{28}) with $n-\lambda^2/n>0$ are poles of the scattering matrix. These poles are located on the negative imaginary axis and, therefore, are associated to anti-bound (virtual) states. We obtain the values for $k'(n)$ replacing  (\ref{28}) in (\ref{27}). The result is

\begin{equation}\label{31}
k'(n)=-\frac{i}{2\alpha} \left( n+\frac{\lambda^2}{n}  \right)\,.
\end{equation}
}

\end{itemize}

In summary, the only existing poles of $\mathcal S(k)$ are those values of $k(n)$ as in (\ref{28}) with negative imaginary part.

\begin{figure}
\centering
\includegraphics[width=0.45\textwidth]{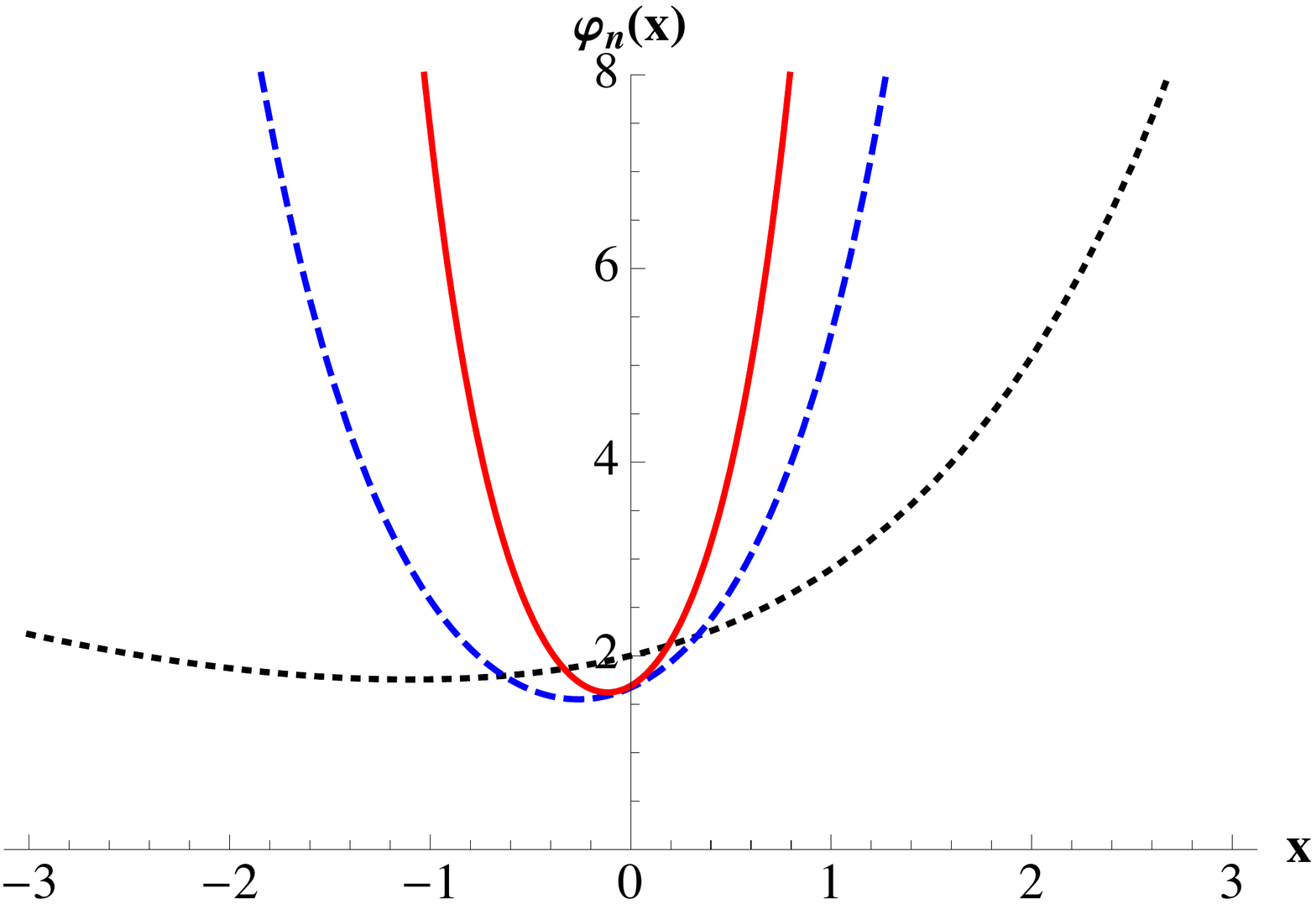}\qquad
\includegraphics[width=0.45\textwidth]{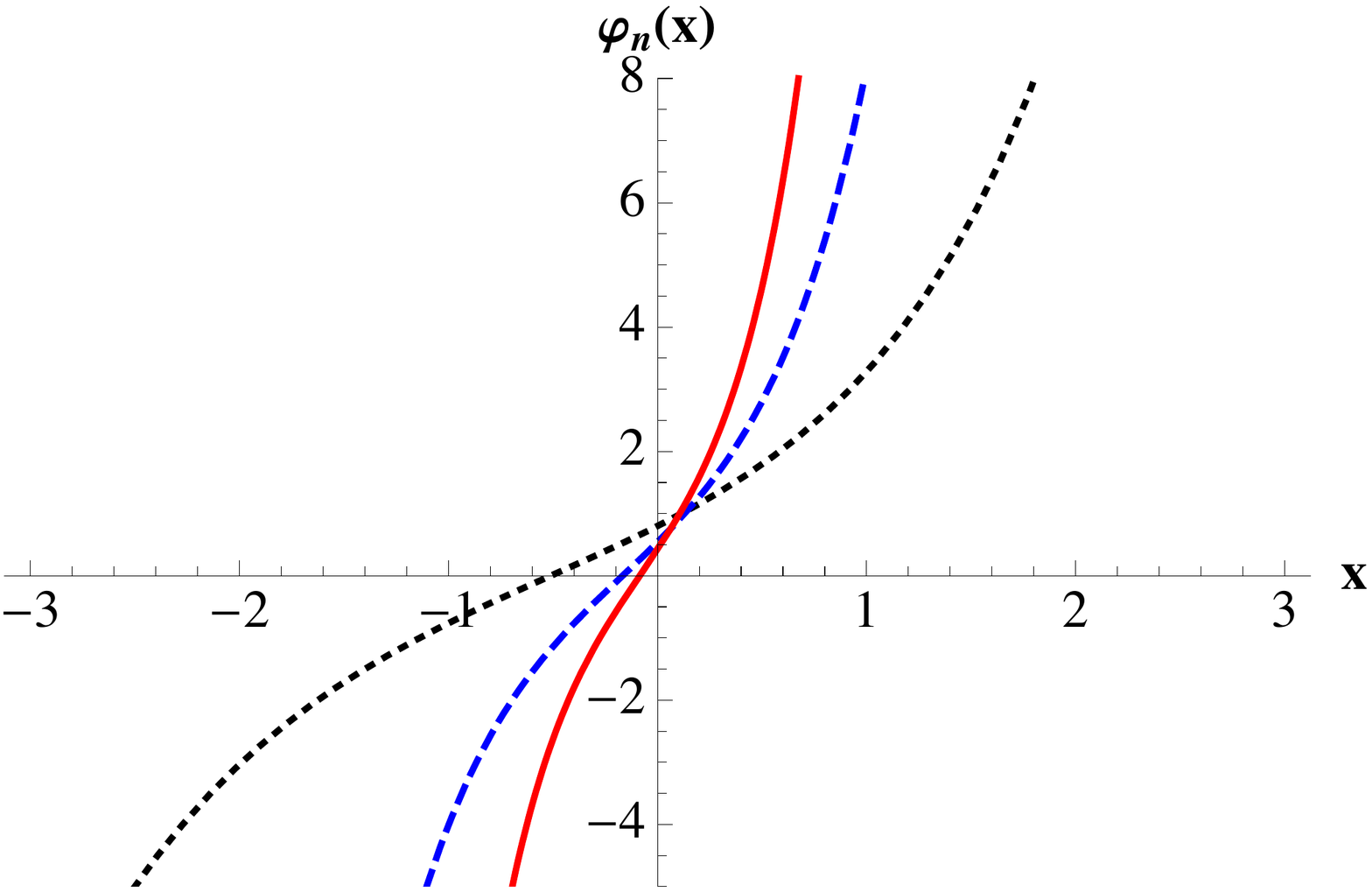}
\caption{\small Wave functions $\varphi_n(x)$ corresponding to anti-bound
states for $V_0=1/2$ and $\alpha=1$. In the figure of the left, we see plots corresponding to the values $n=1$ (dotted), $n=3$ (dashed) and $n=5$ (continuous).  At the figure on the right, we take $n=2$ (dotted), $n=4$ (dashed) and $n=6$ (continuous).}\label{figantibound}
\end{figure}

After these results on the analytic properties of the scattering matrix,  we may summarize our conclusions as follows:

\begin{itemize}

\item{Hamiltonian (\ref{2}) does not have bound states.}

\item{
Resonance behavior is characterized by the existence of pairs of poles located symmetrically with respect to the imaginary axis. Therefore, we must conclude that the hyperbolic step potential does not show resonances. }

\item{Poles of the scattering matrix are given by $k(n)$ as in (\ref{28}) such that Im$k(n)<0$
 They lie on the negative imaginary axis. The same in true for the values $k'(n)$ given by (\ref{31}). These poles should be associated to anti-bound states \cite{NU,BO}. Obviously, the corresponding values of the energy, $E(n)=k^2(n)\hbar^2/2m$, are negative and the wave functions corresponding to these states cannot be normalizable. In Figure \ref{figantibound}, we plot the  wave functions (\ref{20}) for these states which correspond to the values $n=1,\dots,6$. }

\item{ Contrary to a first impression, at the values $k(n)$ with  Im$k(n)>0$, the scattering matrix, or equivalently $T(k)$, has no poles. They are regular points, which show no particular feature. }

\item{The transmission amplitude shows a branch cut with branch points given at $k=\pm \lambda/\alpha$.}

\end{itemize}

Thus far, we have investigated the nature of the singularities of the scattering matrix for the hyperbolic step potential. Next, we study a close relation existing between these
singularities and a kind of supersymmetric partner potentials.  

\section{Supersymmetric partners due to anti-bound states} 

For a matter of convenience and simplicity, we shall choose units such that $\hbar^2/2m=1$ in the sequel.  Whenever convenient, we shall also use the notation $\partial_x:=d/dx$ and $\partial^2_x=d^2/dx^2$. A prime denotes derivation with respect to the variable $x$. 

The point of departure is a Hamiltonian of the form $H=-\partial^2_x+V(x)$, from which we construct a SUSY partner $\widetilde H=-\partial^2_x+\widetilde V(x)$, as follows  \cite{CKS,CHIS}.

Let $E_0$ be the minimum value of the spectrum of $H$. Then, we find a solution of the time independent Schr\"odinger equation $H\psi(x)=\varepsilon\psi(x)$, under the following conditions:

i.) The energy $\varepsilon$ should be lower than $E_0$, i.e., $\varepsilon<E_0$, so that the solution $\psi(x)$ cannot be considered as ``physical'' and is not normalizable. 

ii.) The solution $\psi(x)$ has no zeroes. 

iii.) The inverse of the solution, $1/\psi(x)$ is square integrable, i.e., normalizable.

Next, one determines the function $W(x):=-\psi'(x)/\psi(x)$, which is called the superpotential, and defines the shift operators $A^\pm$ by

\begin{equation}\label{32}
A^\pm:=\mp\partial_x+W(x)\,.
\end{equation}
The Hamiltonian $H$ can be factorized in terms of the shift operators (\ref{32})  as

\begin{equation}\label{33}
H=A^+A^-+\varepsilon=-\partial^2_x+W^2(x)-W'(x)+\varepsilon=-\partial^2_x+V(x)\,.
\end{equation}

Then, we define the supersymmetric partner $\widetilde H$ of $H$ by reversing the
order of the shift operators in the form

\begin{equation}\label{34}
\widetilde H:= A^-A^++\varepsilon= -\partial^2_x+W^2(x)+W'(x)+\varepsilon = -\partial^2_x+\widetilde V(x)\,.
\end{equation}
The relation between $\widetilde V(x)$ and $V(x)$ is given by:

\begin{equation}\label{35}
\widetilde V(x) = V(x) + 2 W'(x)\,.
\end{equation}
We also say that the potential $\widetilde V(x)$ is a SUSY partner of $V(x)$. One important property is that the spectrum of $\widetilde H$ is identical to the spectrum of $H$ with the addition of a bound stated located at $\varepsilon$, which has as wave function $\widetilde \psi(x)=1/\psi(x)$. Other bound or scattering states, $\widetilde \psi(x)$, of $\widetilde H$ are found by applying the shift operator $A^-$ to the wave function $\psi(x)$ corresponding to the bound or scattering state of $H$ with the same energy, i.e., 

\begin{equation}\label{36}
A^-:\psi(x) \to \widetilde \psi(x)=A^-\psi(x)\,.
\end{equation}

This process can be iterated in order to get second,  or higher, order SUSY partner potentials. This is named $n$-SUSY or Darboux-Crum transformation \cite{FLISHFLASH,MO}. For the second order, the point of departure are two eigenfunctions, $\psi_1(x)$ and $\psi_2(x)$, of $H$ with respective eigenvalues $E_0>\varepsilon_1>\varepsilon_2$ under the condition that the Wronskian ${w}(\psi_{1},\psi_{2})=\psi_1(x)\,\psi'_2(x) - \psi'_1(x)\,\psi_2(x)$ does not vanish. Then, one defines a superpotential $\widetilde W(x)$ as

\begin{equation}\label{37}
{\widetilde W}(x)=-\dfrac{w'(\psi_{1},\psi_{2})}{w(\psi_{1},\psi_{2})}\,.
\end{equation}
The second order partner potential is

\begin{equation}\label{38}
\widetilde{\widetilde V}(x) = V(x) + 2 \widetilde W'(x)\,.
\end{equation}

We may go on with this process so as to obtain a sequence of Hamiltonians and potentials, the $n$-th being the $n$-th SUSY partner of the original Hamiltonian, $H$, and potential, $V$, respectively.

\subsection{SUSY partners with anti-bound states}

The anti-bound states of the hyperbolic step potential are particularly well
suited to apply this method to produce supersymmetric partner Hamiltonians. In fact,  they lead to a shape invariant hierarchy of potentials.

Let us construct an explicit example. Take $n=1$, $V_0=1/2$ and $\alpha=1$. The energy for the first anti-bound state is $\varepsilon_1=E(1)=k^2(1)=-0.0625$ and its wave function is

\begin{equation}\label{39}
\varphi_{1}(x)=(1+e^x) \,e^{-x/4}\,.
\end{equation} 
Due to the presence of exponentials in (\ref{39}), this function does not vanish at any point. Furthermore, its inverse $\widetilde \varphi_1(x)=1/\varphi_1(x)$ is normalizable and is a bound state of the SUSY partner, $\widetilde H$, of $H$ with energy $E(1)$. This partner has the form $\widetilde H=-\partial^2_x+\widetilde V(x)$ with 

\begin{equation}\label{40}
\widetilde V(x)=
\frac{1}{4}\,\left(1+\tanh\frac{x}{2}\right)-\frac{1}{2} \, {\rm sech}^2{\frac{x}{2}}\,.
\end{equation}
This partner potential $\widetilde V(x)$  is known as the Rosen Morse II potential \cite{CKS}. 
It is noteworthy that, although the hyperbolic step potential does not have bound states, its first SUSY partner has just one.

\begin{figure}
\centering
\includegraphics[width=0.45\textwidth]{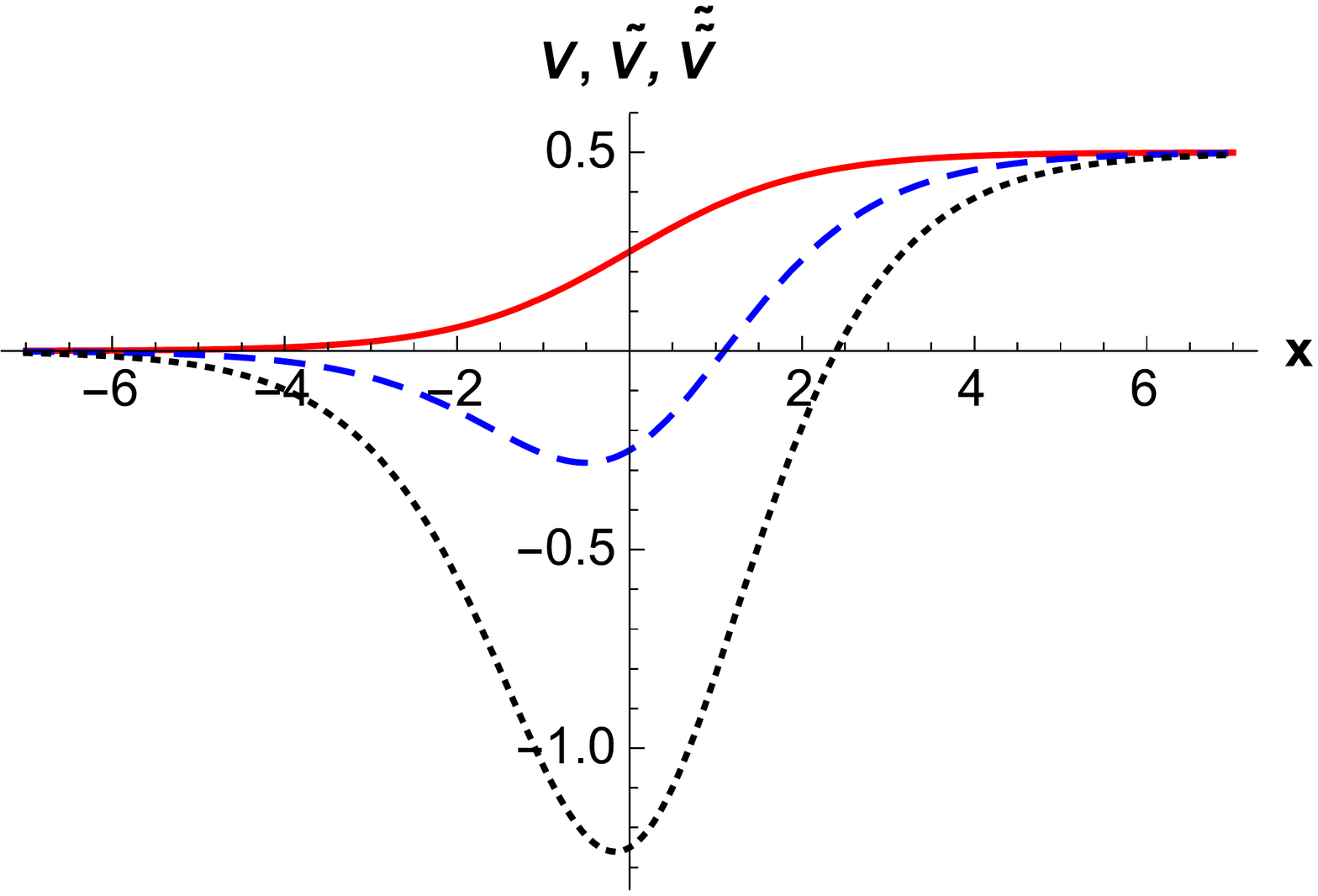}\qquad
\includegraphics[width=0.45\textwidth]{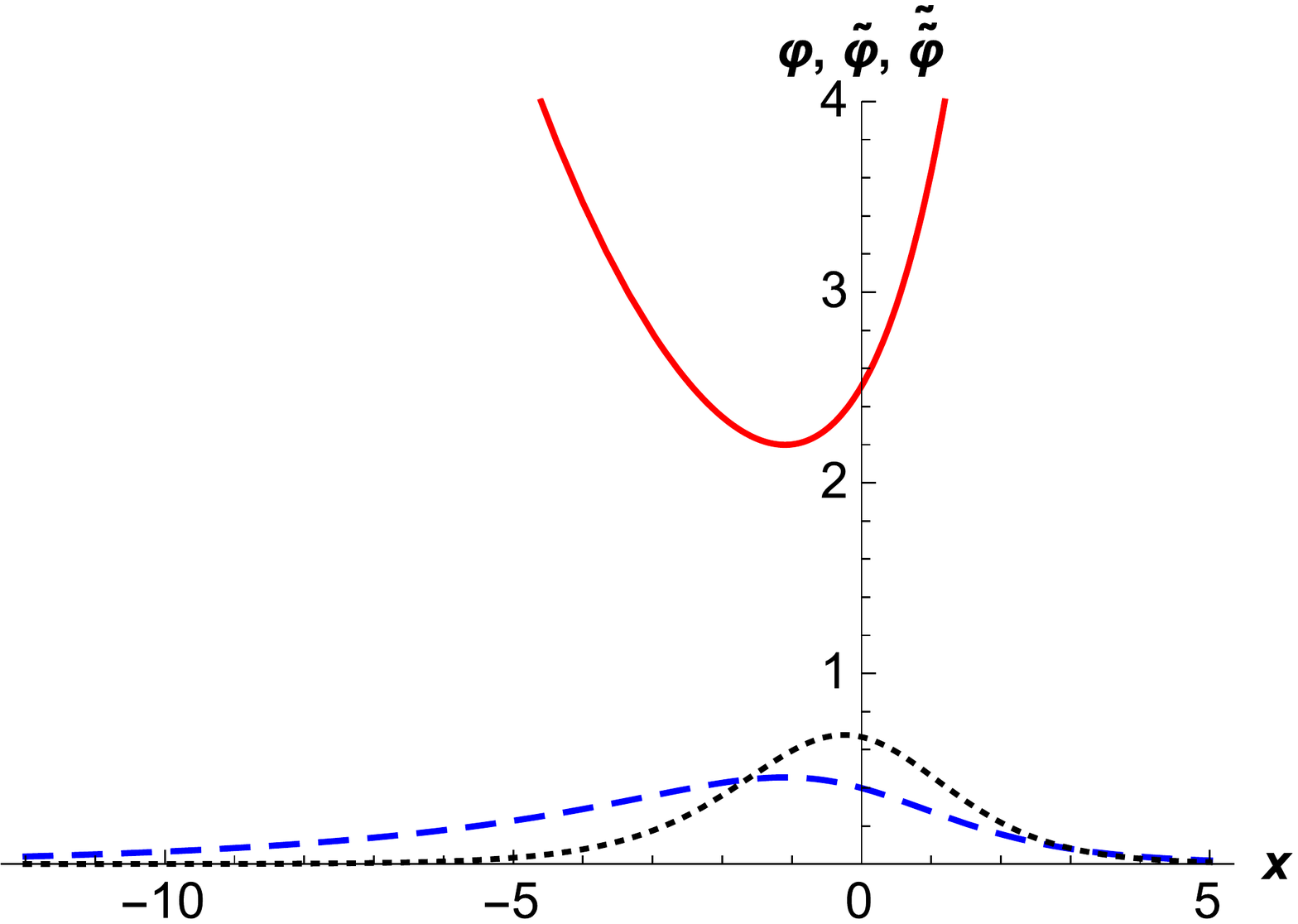}
\caption{\small Left: Comparison of the hyperbolic step potential (continuous line) with its first order  (\ref{38}) (dashed) and  second order  SUSY partner potentials (\ref{40}) (dotted). Right: Wave function (\ref{37}) for the anti-bound
state $\varphi_1(x)$   (continuous line) and  wave
functions for the ground state $\tilde{\varphi}_1(x)=1/\varphi_1(x)$
(dashed) and second bound state $\tilde{\tilde{\varphi}}_2(x)$ (dotted) of $\widetilde{\widetilde H}$.}\label{fig_step_partner}
\end{figure}

We may proceed further and obtain higher order SUSY transformations generated by the anti-bound states of the hyperbolic step potential, so as to find the shape invariant hierarchy of Rosen Morse II potentials (see Figure \ref{fig_step_partner}). 
For example, let us choose  the first two anti-bound states of the hyperbolic step potential for $V_0=1/2$, where the wave functions 
$\varphi_1(x)$ is given by (\ref{37}) and $\varphi_2(x)$, according to (\ref{20}), is

\begin{equation}\label{41}
\varphi_{2}(x)= {(1+e^x)(-3+5\,e^x)}{ e^{-7x/8}}\,.
\end{equation} 
Furthermore, the Wronskian $w(\varphi_1,\varphi_2)$ is different from zero at all real points. Using the above procedure, we can construct the second SUSY potential, which is

\begin{equation}\label{42}
\widetilde{\widetilde V}(x)=\frac{1}{4}\,\left(1+\tanh\frac{x}{2}\right)-\frac{3}{2}\, {\rm sech}^2{\frac{x}{2}}\,.
\end{equation}
The Hamiltonian $\widetilde{\widetilde H}:=-\partial^2_x+\widetilde{\widetilde V}(x)$ has two bound states with energies $\varepsilon_1$ and $\varepsilon_2$. 

In Figure  \ref{fig_step_partner}, at the left, we compare the hyperbolic step potential with its first and second SUSY partners. At the right, we compare $\varphi_1(x)$ with wave functions of the ground states of the first and the second SUSY partner Hamiltonians ${\widetilde H}$ and $\widetilde{\widetilde H}$. 

Finally, we may go on with the procedure and obtain third, forth order SUSY partners and so on. The $n$-th potential partner is 

\begin{equation}\label{43}
{\widetilde V}^{(n)}=\frac{1}{4}\,\left(1+\tanh\frac{x}{2}\right)-\frac{n}{4}(n+1)\, {\rm sech}^2{\frac{x}{2}}\,, \qquad n=1,2,\dots\,,
\end{equation}
which is the family of shape invariant Rosen Morse II potentials studied in \cite{CKS}. Hamiltonians with potentials as in (\ref{43}) have exactly $n$ bound states with energies 
$\varepsilon_s=E(s)=k^2(s)$, $s=1,2,d\dots,n$. 
The explicit form of their respective wave functions is  calculated in a similar way \cite{CKS,FLISHFLASH}. 

\section{An analysis on time delay}

Let us split this discussion on time delay into two parts: quantum and classical time delay. In the quantum calculations along this section consider $\hbar^2/2m=1$, so that $k=\sqrt E$, and $\alpha=1$.  

\subsection{Quantum time delay}

Let us assume that a wave packet evolves freely until it reaches an interacting potential. Both reflected and transmitted wave packets undergo a time delay with respect to the time elapsed during the free motion. First, let us consider the phases of the amplitudes $r$, and $t$,

\begin{equation}\label{44}
r= |r|e^{i\delta_r},\qquad t= |t| e^{i\delta_t}\,.
\end{equation}
Then, the reflection or transmission time delay may be estimated by the derivative of the phase shift for the reflection (transmission) amplitude $\delta_r$ ($\delta_t$).
This is the Wigner time delay $\tau$ \cite{W,S,T,DB,BO,GP} and is defined as

\begin{equation}\label{45}
\tau=\frac{1}{k}\,\frac{d\,\delta}{d\,k}\,.
\end{equation}

The objective of the present subsection is to evaluate the effect of a SUSY transformation on the reflection and transmission amplitudes. To this end, let us go back to the situation described in (\ref{3a}). Then, we perform a SUSY transformation using the anti-bound state wave function $\varphi_n(x)$ of the hyperbolic step potential, with $n$ odd. In the sequel, $\psi(x)$ is the wave function with asymptotic values given by (\ref{21}), $W_n(x)=-\varphi_n'(x)/\varphi_n(x)$ is the superpotential and  $A_n^-$ is the shift operator with minus sign as in (\ref{32}). If we want to obtain the SUSY partner of $\psi(x)$, then  we use the definition in (\ref{36}) and apply $A_n^-$:

\begin{equation}\label{46}
\widetilde \psi(x):= A_n^-\,\psi(x)=\left( \frac{d}{dx} +W_n(x)    \right) \psi(x)\,.
\end{equation}

However,  we really need  the asymptotic expressions of $\psi(x)$, which may be obtained with the asymptotic values, $W_n^\pm(x)$, of the superpotential $W_n(x)$. In order to find  $W_n^\pm(x)$, we need to know the asymptotic values of $\varphi_n(x)$ first. For the regions $x\to\pm\infty$, these are, respectively,

\begin{equation}\label{47}
\varphi^+_n(x)=e^{\frac{1}{2}(n+\frac{V_0}{n}) x}\,,\qquad \varphi^-_n(x)=e^{-\frac{1}{2}(n-\frac{V_0}{n}) x}\,.
\end{equation}
Then, it is natural to define $W_n^\pm(x):=-{\varphi'}_n^\pm(x) /\varphi_n^\pm(x)$, so as to obtain:

\begin{equation}\label{48}
W_n^+=-\frac{1}{2}\,\left(n+\frac{V_0}{n}\right)\,,\qquad W_n^-=\frac{1}{2}\,\left(n-\frac{V_0}{n}\right)\,.
\end{equation}
In consequence, the asymptotic values of (\ref{46}) as $x\to\pm\infty$ are given by

\begin{equation}\label{49}
\widetilde \psi_n^-(x)= \left (\frac{d}{dx}+W_n^-\right)(e^{ikx}+r(k)\,e^{-ikx})\,,\qquad \widetilde \psi_n^+(x) = \left (\frac{d}{dx}+W_n^+\right)\,t(k) \,e^{ik'x}\,.
\end{equation}
We perform the derivatives in (\ref{49}) and divide both expressions by $ik+W_n^-$ in order to maintain $e^{ikx}$ as incident plane wave from the left. 
As a consequence, the new reflection, $\widetilde r(k)$, and transmission, $\widetilde t(k)$, amplitudes for the first SUSY partner 
are given by 

\begin{equation}\label{50}
\widetilde r= r \,\left(\dfrac{-i k+W_n^-}{i k+W_n^-}\right)=r\, e^{i \Delta_r}\,,
\qquad 
{\tilde t}= t \,\left(\dfrac{i k'+W_n^+}
{i k +W_n^-}\right)=t\, e^{i \Delta_t}\,.
\end{equation}
Relations (\ref{50}) show that $|\tilde r| = | r|$ and $|\tilde t| = | t|$, so both reflection, $R(k)$, and transmission, $T(k)$, coefficients do not change after these SUSY transformations \cite{CKS,FLISHFLASH}. Nevertheless, the phases of the amplitudes undergo a change. This is the crucial point to compare time delays for the partner potentials. In fact, if $\delta_{\widetilde r}$ and $\delta_{\widetilde t}$ are the phases of $\widetilde r$ and $\widetilde t$, respectively, we have

\begin{equation}\label{51}
\delta_{\widetilde r} = \delta_r + \Delta_r\,,\qquad
\delta_{\widetilde t} = \delta_t + \Delta_t\,.
\end{equation} 
Phase differences given by the arguments $\Delta_r$ and $\Delta_t$ are  evaluated from (\ref{50}) as

\begin{equation}\label{52}
\Delta_r=-2 \arctan{\frac{k}{ W_n^-}}\,,\qquad 
\Delta_t=\dfrac{\pi}{2}+ \arctan{\frac{-W_n^+}{k'}}-\arctan{\frac{k}{W_n^-}}\,.
\end{equation}

The corresponding relations for the Wigner time delay are obtained by differentiation of (\ref{51}),
with respect to $k$ as in (\ref{45}):

\begin{equation}\label{53}
\tau_{\tilde r} = \tau_r + (\Delta\tau)_r\,,\qquad
\tau_{\tilde t} = \tau_t + (\Delta\tau)_t\,.
\end{equation}
The explicit forms of the time delay differences $(\Delta\tau)_r$ and $(\Delta\tau)_t$ are 

\begin{equation}\label{54}
(\Delta\tau)_r=-\frac{2 W_n^-}{k^3+k\,(W_n^-)^2}\,,\quad 
(\Delta\tau)_t=-\frac{W_n^-}{k^3+k\,(W_n^-)^2}+\frac{W_n^+}{\sqrt{k^2-V_0}\,(k^2+(W_n^+)^2-V_0)}\,.
\end{equation}
 Similarly, we find phases $\delta_{ \tilde{\tilde r} }$, $\delta_{ \tilde{\tilde t} }$ and Wigner 
times $\tau_{\tilde{\tilde r}}$, $\tau_{\tilde{\tilde t}}$ for all  second order partner potentials $\widetilde{\widetilde V}$.  Needless to say that we use $k'=\sqrt{k^2-V_0}$ in all operations.
 
\begin{figure}
\centering
\includegraphics[width=0.45\textwidth]{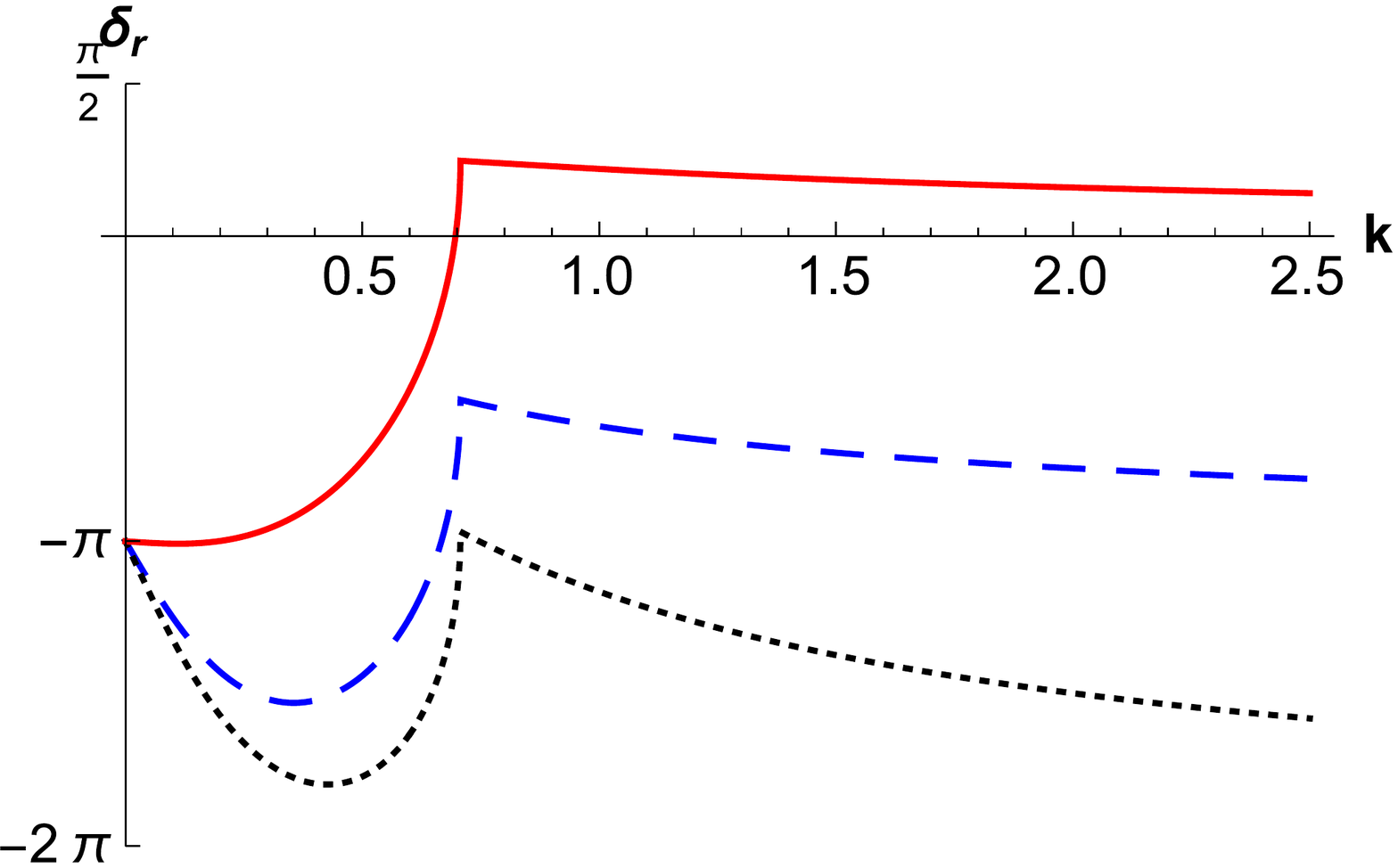}\qquad
\includegraphics[width=0.45\textwidth]{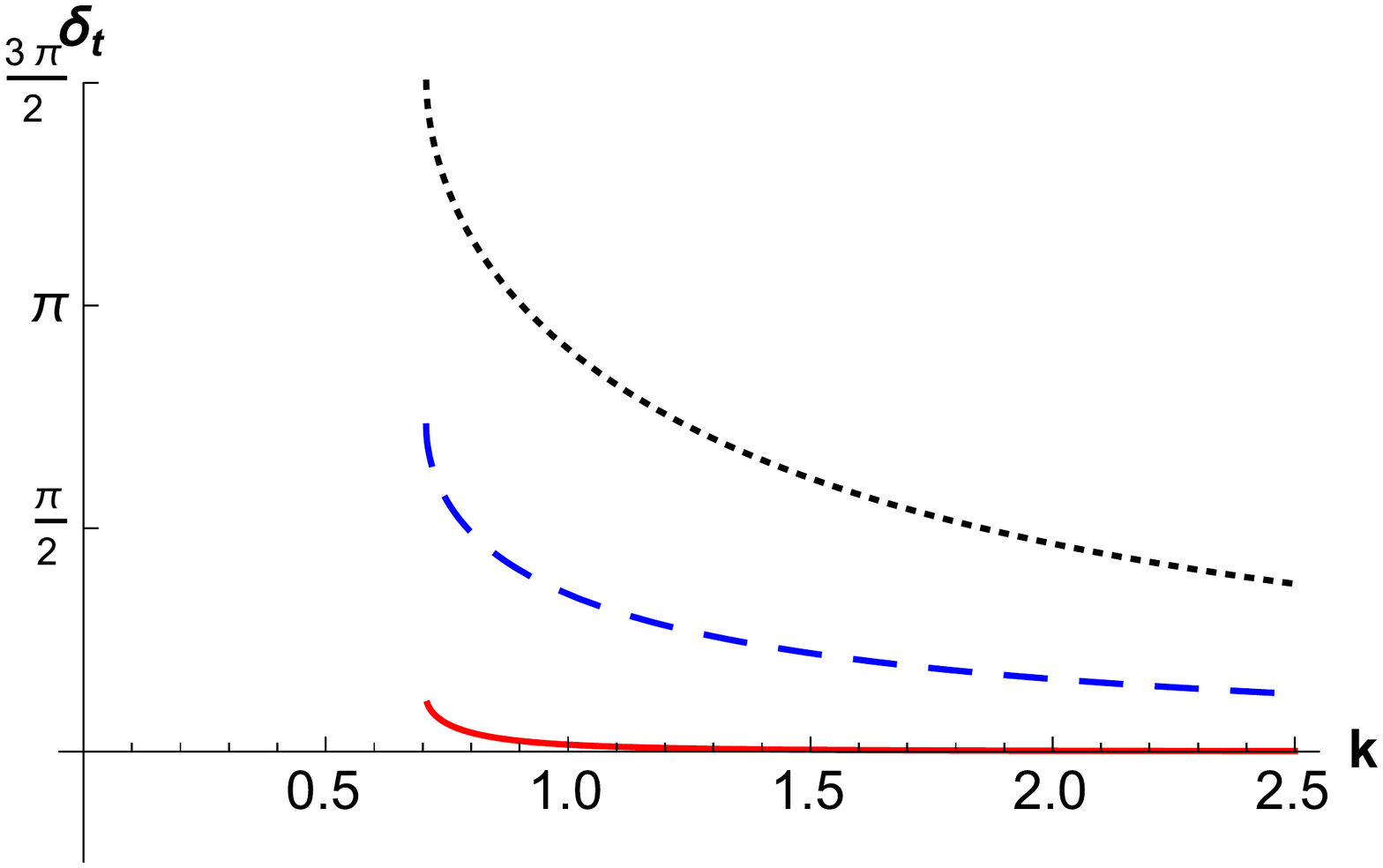}
\caption{\small Variations with respect to $k$ of the phases  $\delta_r, \delta_t$ (continuous red line) for the hyperbolic step potential, 
$\delta_{\tilde r},  \delta_{\tilde t}$ (dashed blue line) for first order SUSY partner
and  $\delta_{\tilde{\tilde r}},  \delta_{\tilde{\tilde t}}$ (dotted black line) for second order SUSY partner of the hyperbolic step potential. }\label{phase_tr}
\end{figure}

\begin{figure}
\centering
\includegraphics[width=0.45\textwidth]{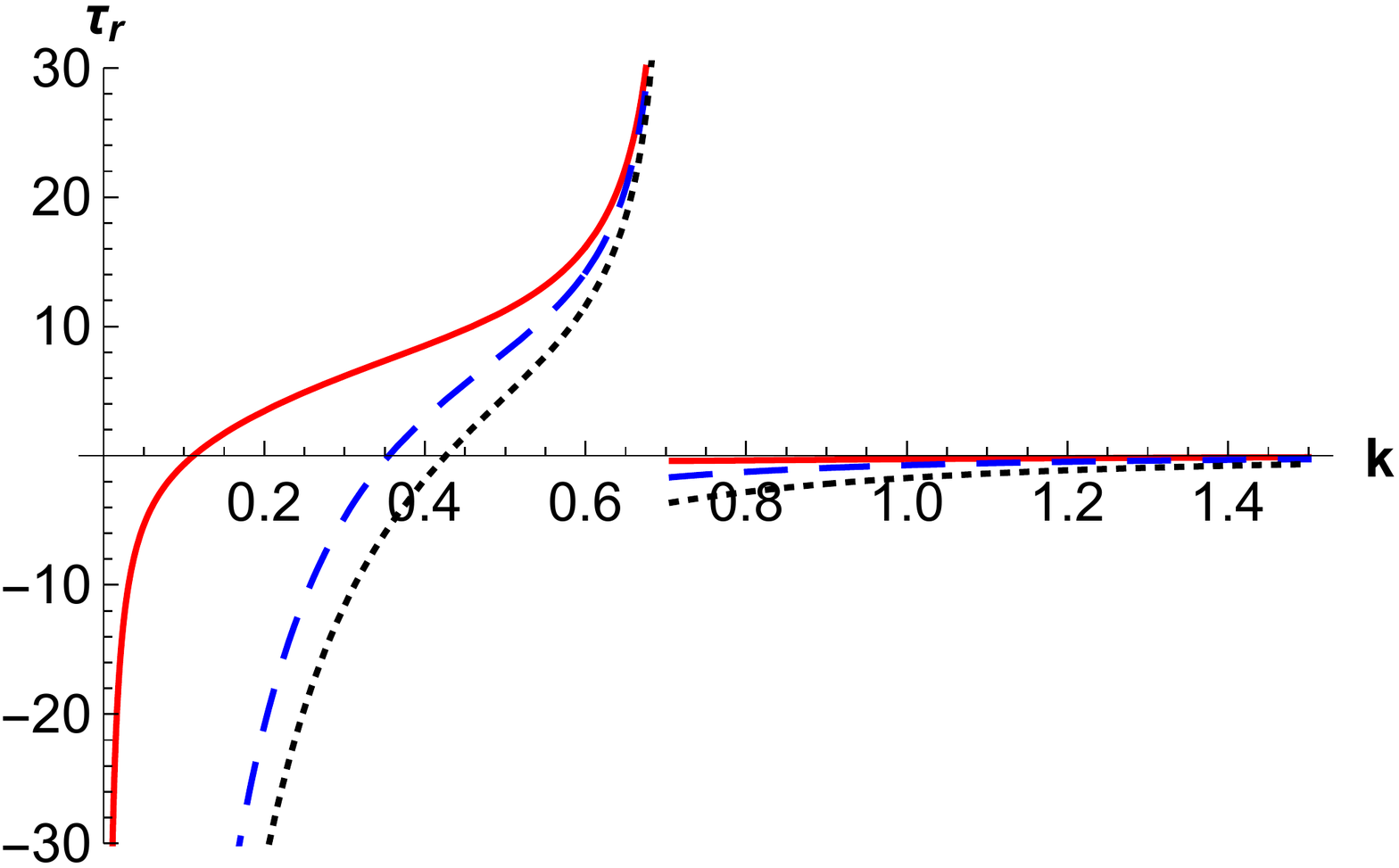}\qquad
\includegraphics[width=0.45\textwidth]{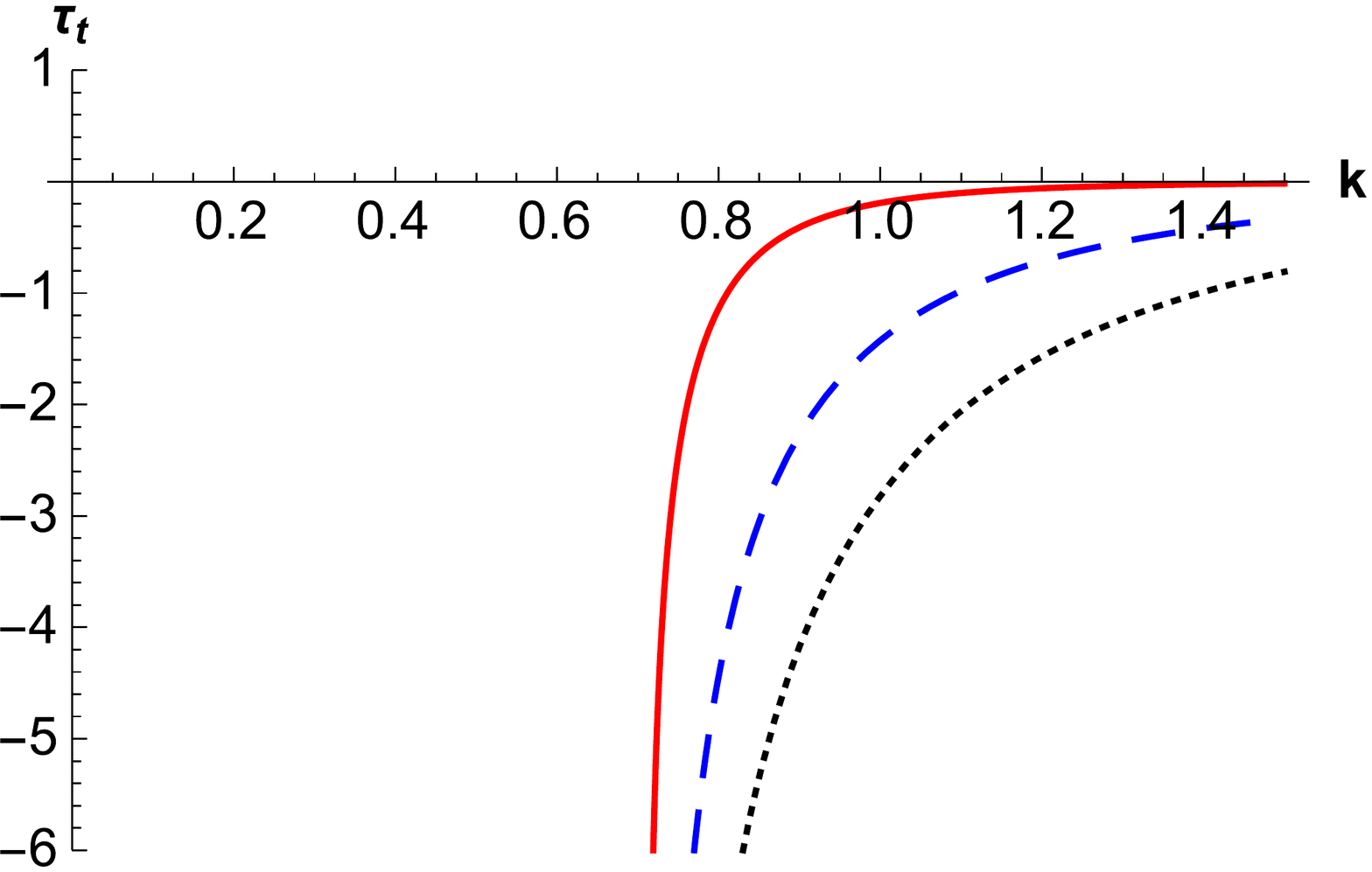}
\caption{\small Wigner time delay   $\tau_r, \tau_t$  (continuous line) for the hyperbolic step potential; $\tau_{\tilde r},  \tau_{\tilde t}$ (dashed line)   the first order partner $\widetilde V$ and $\tau_{\tilde{\tilde r}},  \tau_{\tilde{\tilde t}}$ (dotted line) and  the second order partner $\widetilde{\widetilde V}$. }\label{delay_tr}
\end{figure}

Phases and Wigner time delays for the hyperbolic step potential and its first and second order SUSY Rosen Morse II partners, constructed as above, can be obtained analytically, through often cumbersome calculations.   Their plots are shown in Figures \ref{phase_tr} and \ref{delay_tr}. In all our graphics, we have chosen $V_0=1/2$ and $n=1$. 

In Figure \ref{phase_tr}, we plot the phase shifts corresponding to the reflection and transmission amplitudes for the hyperbolic step potential and its first and second order partners. At the left, we see that the reflection phase shift from $k=0$ to $k=+\infty$ changes by $\pi$ for the step potential, by $0$ for its first order partner and by $-\pi$ for its second order partner. This is in agreement with the arguments posed in \cite{W}, according to which each bound state adds a $-\pi$ to the total phase.  As we observe from the left graphic of Figure 7, the behavior of the reflection phase shift undergoes a sudden change near $V_0$. 
This direction change at $E=V_0$ implies a discontinuity of the delay time, as shown in Figure \ref{delay_tr}.
In the transmission case 
(see Figure \ref{phase_tr}, right), where the energy of the incident plane wave is greater than $V_0$, the phase is higher for the partner potentials with bound states than
for the step potential, but they have the same asymptotic behavior at $k\to +\infty$.

In Figure \ref{delay_tr}, we represent Wigner time delays in terms of $k$ for the same potentials. In both cases, reflection (left) and transmission (right), the more number of bound states the potential has the smaller is the time delay. Reflection time delay for $E<V_0$ may be either positive or negative, when $\varepsilon$ approaches to $V_0$ the time delay goes to infinity and right after $V_0$ is small and goes to zero as $E$ goes to infinity. Transmission delay time is always negative and also goes to zero as $E$ goes to infinity. The presence of a negative transmission time delay has been reported by Hartman for square barriers \cite{HAR}.

\subsection{Classical time delay}

We may also compute the classical time delay for the hyperbolic step potential and provide an explicit expression for it. From the fact that the total energy $E=\frac12 mv^2(x)+V(x)$ is a constant of motion, we obtain the velocity as

\begin{equation}\label{55}
v(x)=\sqrt{\frac{2(E-V(x))}{m}}\,.
\end{equation}
Then, the time for a classical particle to go between two points
at $x=d_1$ and $x=d_2$ is obtained by

\begin{equation}\label{56}
T =\int_{d_1}^{d_2}{\dfrac{dx}{v(x)}}=\int_{d_1}^{d_2}{\frac{dx}{\sqrt{\dfrac{2(E-V(x))}{m}}}}\,.
\end{equation}

Let $d$ be a distance from the origin such that the value of the potential at $-d$ be negligible. The time it takes a free particle with energy $E$ and speed $v=\sqrt{2E/m}$ to cross the interval $[-d,d]$ is $T_{\rm free}=2d/v$.  Assume that $0<E<V_0$, then the classical particle with this energy bounces back at a turning point, $x_{\rm turn}$, defined by the equation $E=V(x_{\rm turn})$. Therefore, we may compare the difference of times between the free motion and the time $T$ that the particle under the action of hyperbolic step potential elapses between $d_1=-d$ and $d_2= x_{\rm turn}$ and back. This is given by the following expression

\begin{equation}\label{57}
T =\int_{d_1}^{d_2}{\dfrac{dx}{v(x)}}=2 \int_{-d}^{x_{\rm turn}}{\frac{dx}{\sqrt{\dfrac{2(E-V(x))}{m}}}}\,.
\end{equation}
Thus, we define the time delay, $\tau_r^c$ for the reflected particle as

\begin{equation}\label{58}
\tau_{r}^c={\it T}-{\it T}_{\rm free}\,,
\end{equation}
and taking $V_0=1$ and $2m=1$, for this case, it has the form

\begin{equation}\label{59}
\tau_{r}^c={\it T}(-d\to x_{\rm turn})+{\it T} (x_{\rm turn}\to -d)-\frac{d}{k}\,. 
\end{equation}
Expressions like  $T(a\to b)$ mean the time used by a classical particle under the action of the hyperbolic step potential to go from the point $x=a$ to the point $x=b$. Here, $k=\sqrt E$. 

If $E>V_0$, it makes sense to obtain the time delay for the transmitted particle. In this case, we choose $d_1=-d$ and $d_2=d$ in (\ref{57}) to obtain a crossing time $T$. Again, the transmission delay time is defined as $\tau_t^c:=T-T_{\rm free}$ and the result is

\begin{equation}\label{60}
 \tau_{t}^c={\it T} (-d\to d)-\frac{d}{2\sqrt{k^2-1/2}}-\frac{d}{2k}\,. 
\end{equation}

The primitive of the function under the integral sign in (\ref{57}) is given by

\begin{equation}\label{61}
T(x)=\int \frac{dx}{\sqrt{\frac{2(E-V(x))}{m}}} = \left( -\frac{{\rm arctanh}{\frac{\sqrt{-1+4k^2-\tanh{x/2}}}{2k}}}{k}
+\frac{{\rm arctanh}{\frac{\sqrt{-1+4k^2-\tanh{x/2}}}{\sqrt{2-4k^2}}}}{\sqrt{\frac12-k^2}}\right)\,.
\end{equation}

\begin{figure}
\centering
\includegraphics[width=0.45\textwidth]{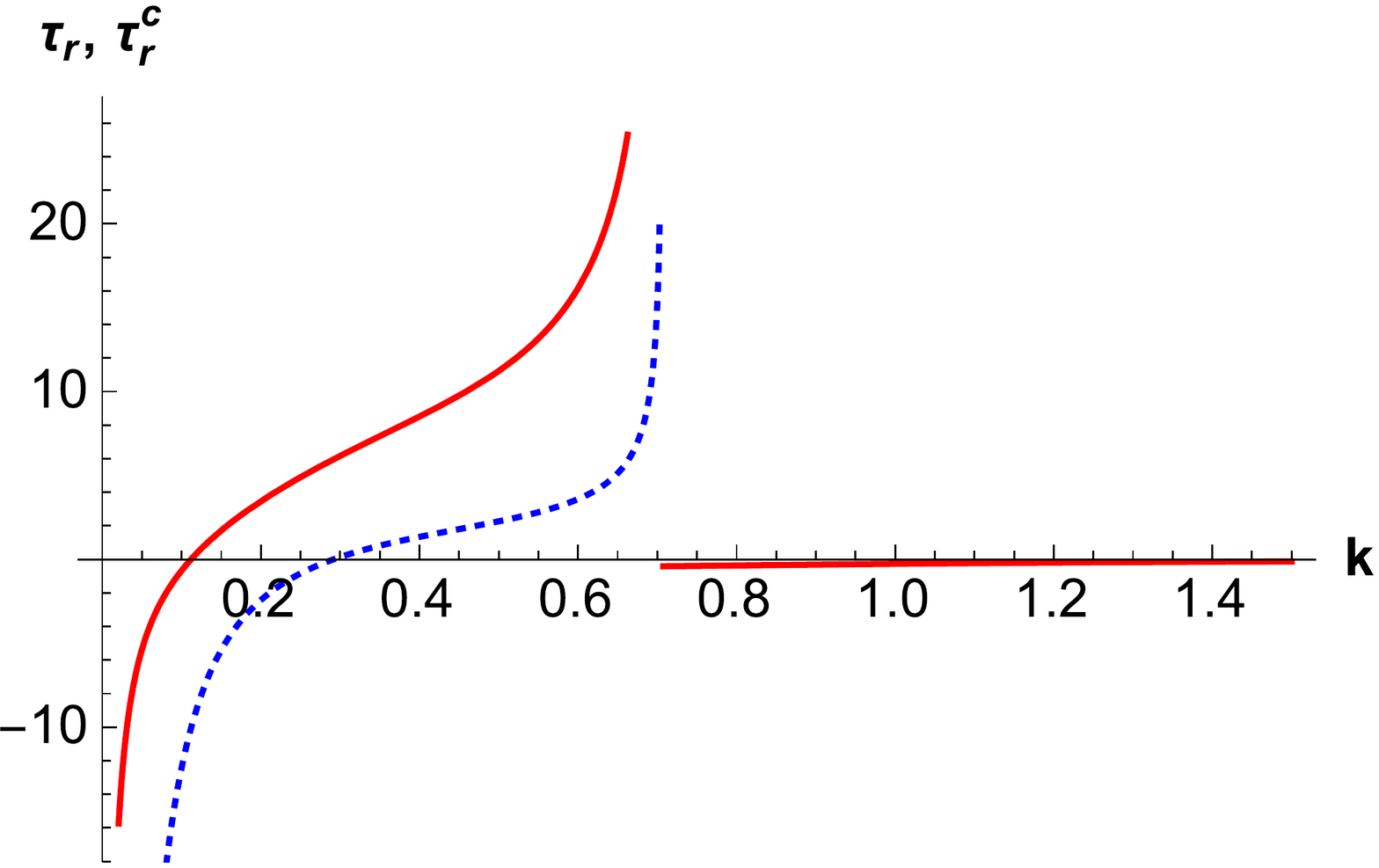}\qquad
\includegraphics[width=0.45\textwidth]{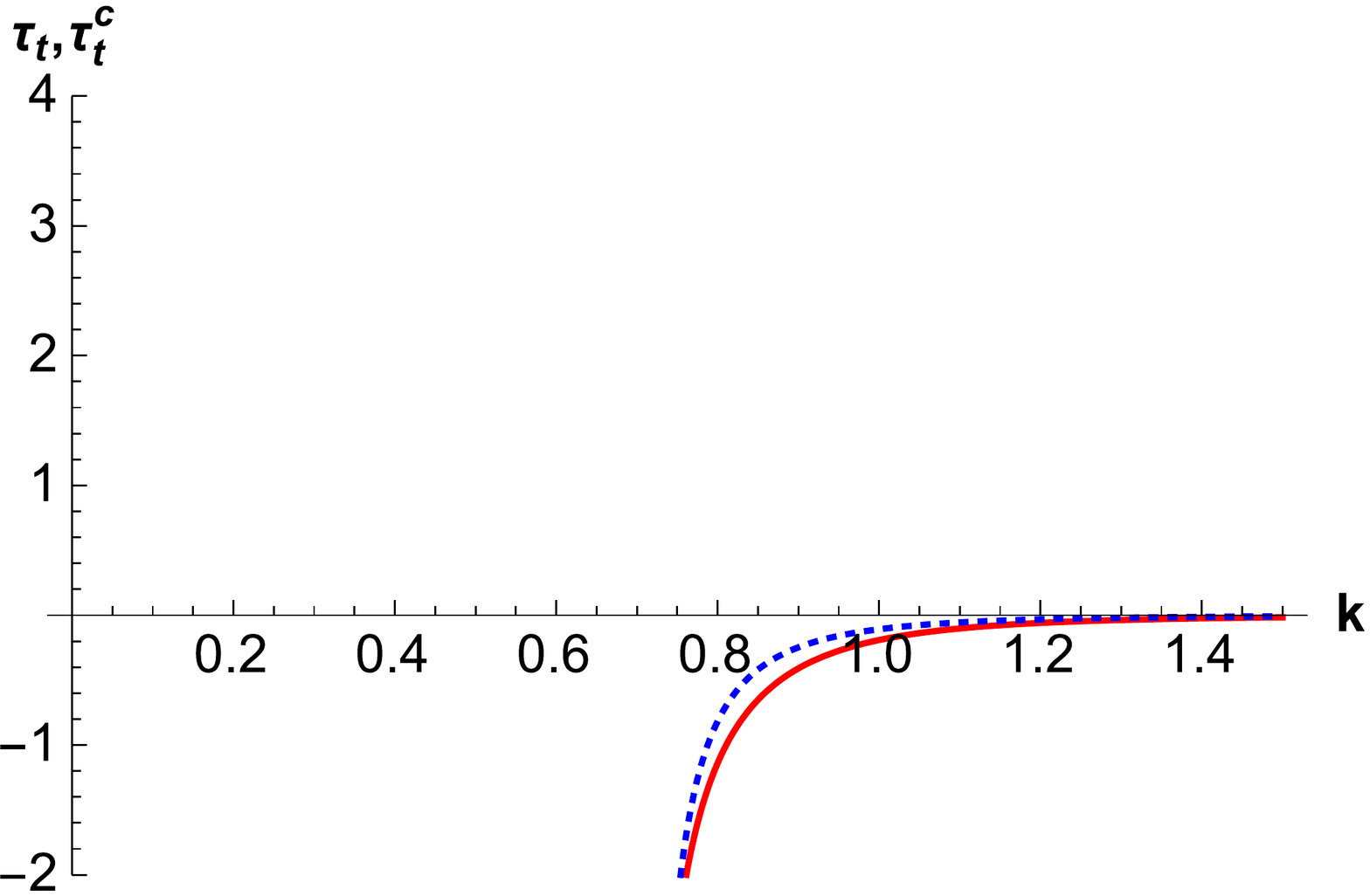}
\caption{\small Classical (dotted line) and quantum (continuous line) time delays for the hyperbolic step potential. }\label{delayclassical}
\end{figure}
With the aid of (\ref{61}), we can obtain the delay times (\ref{59}) and (\ref{60}). In Figure \ref{delayclassical}, we have compared classical and quantum delay times. The similarity of both is surprising, including their behavior at the limits $E\to 0$ and $E\to+\infty$.  

\section{Concluding remarks}

One dimensional exactly solvable models give relevant information in quantum mechanics. In particular, there are not many models that give exact solutions for resonances, anti-bound states and other features of scattering processes. For most of known models resonance or anti-bound (virtual) poles are determined approximately through transcendental equations. 

One among these models is the one dimensional Hamiltonian $H=-\partial^2_x+V(x)$, where $V(x)$ is the hyperbolic step potential. In addition, this potential shows properties that make it particularly interesting. 

First of all, $H$ does not have bound states, so that all its interest lies in its scattering properties. The most important of these properties are: 

i.) The scattering matrix in momentum representation, ${\cal S}(k)$, is not unitary, although it obeys some sort of {\it modified} unitarity as shown in (\ref{17}). 

ii.) The scattering matrix ${\cal S}(k)$ depends on $k$ through an square root,
$k'(k)=\sqrt{k^2-\lambda^2/\alpha^2}$, which implies that the analytic continuation will have a branch cut.
Besides this, it shows an infinite number of simple poles on the negative imaginary axis. These poles are  an evidence of the presence of anti-bound states.

The exact solvability has other advantages. In particular, we may use the wave functions of the anti-bound states in order to construct  SUSY partners of the hyperbolic step potential, one  for each anti-bound pole. We have shown that these partner potentials coincide with the series of Rosen Morse II potentials.

We have computed the reflection and transmission Wigner time delays for the hyperbolic step potential and compared them with its SUSY partners. We have arrived to the following conclusions: 

i.) Time delays are larger for the hyperbolic step potential than for any other of its SUSY partners. The higher the number of bound states a partner has, the shorter the time delay is.

ii.) Time delays have a singular behavior near the potential height $V_0$. 

iii.) Time delays go to zero for very high energies.

iv.) If we compare between time delays for the quantum and classical hyperbolic step potentials, we show a remarkable similarity among them.

In the near future we plan to report on other hypergeometric asymmetric potentials
including Eckart (corresponding to the Coulomb potential in a hyperboloid) and some other types of Rosen-Morse (characterized by having a barrier shape).

\section*{Acknowledgements}

We acknowledge partial financial support to  the Spanish MINECO (Project MTM2014-57129-C2-1-P), Junta de Castilla y Le\'on (VA057U16) and GIR of Mathematical Physics at UVa.
 \c{S}.~Kuru acknowledges Ankara University and the
warm hospitality at the Universidad de Valladolid, Dept.\
de F\'isica Te\'orica,  where a part of this work has been done.

\end{document}